\documentclass[american,aip, notitlepage, groupedaddress, superscriptaddress, twocolumn, reprint]{revtex4-1}

\usepackage[LGR,T1]{fontenc}
\usepackage[latin9]{inputenc}
\usepackage{babel}
\usepackage{amstext}
\usepackage{amssymb}
\usepackage{stmaryrd}
\usepackage{graphicx}
\usepackage{subscript}
\usepackage[unicode=true,pdfusetitle,
 bookmarks=true,bookmarksnumbered=false,bookmarksopen=false,
 breaklinks=false,pdfborder={0 0 1},backref=false,colorlinks=false]
 {hyperref}

\makeatletter

\DeclareRobustCommand{\greektext}{%
  \fontencoding{LGR}\selectfont\def\encodingdefault{LGR}}
\DeclareRobustCommand{\textgreek}[1]{\leavevmode{\greektext #1}}
\ProvideTextCommand{\~}{LGR}[1]{\char126#1}

\providecommand{\tabularnewline}{\\}

\usepackage{hyperref}
\date{\today}

\makeatother

\begin{document}
\title{Granular superconductors for high kinetic inductance and low loss
quantum devices }
\author{Aviv Glezer Moshe}\thanks{Electronic mail: avivmoshe@mail.tau.ac.il}
\affiliation{Raymond and Beverly Sackler School of Physics and Astronomy, Tel Aviv
University, Tel Aviv 69978, Israel}
\affiliation{Department of Physics and Department of Electrical and Electronic
Engineering, Ariel University, P.O.B. 3, Ariel 40700, Israel.}
\author{Eli Farber}
\affiliation{Department of Physics and Department of Electrical and Electronic
Engineering, Ariel University, P.O.B. 3, Ariel 40700, Israel.}
\author{Guy Deutscher}
\affiliation{Raymond and Beverly Sackler School of Physics and Astronomy, Tel Aviv
University, Tel Aviv 69978, Israel}

\begin{abstract}
Granular aluminum is a promising material for high kinetic inductance
devices such as qubit circuits. It has the advantage over atomically
disordered materials such as NbN\textsubscript{x}, to maintain a
high kinetic inductance concomitantly with a high quality factor.
We show that high quality nano-scale granular aluminum films having
a sharp superconducting transition with normal state resistivity values
of the order of $1\times10^{5}\,\mu\Omega\,cm$ and kinetic inductance
values of the order of $10\,nH/\oblong$ can be obtained, surpassing
state of the art values. We argue that this is a result of the different
nature of the metal-to-insulator transition, being electronic correlations
driven (Mott type) in the former and disorder driven (Anderson type)
in the latter.
\end{abstract}

\maketitle

In recent years superconducting (SC) qubit circuits (QC) have evolved
considerably. The first type, the so-called Cooper pair box \citep{Bouchiat1998,Nakamura1999},
was basically composed of an island connected through a small Josephson
junction (JJ) to a large pair reservoir, while a voltage V could be
applied to it through a gate. States differing by one pair on the
island could be weakly coupled through the JJ, creating a two-level
system. More recently a different type of qubit, consisting of a small
JJ shunted by a high inductance element, was developed to eliminate
charge related noise and de-coherence. The large inductance element
consisted originally of a one-dimensional array of large size JJ \citep{Mooij1999,Manucharyan2009}.
More recently it was found that it could simply consist of a narrow
line of a highly disordered superconductor such as NbN\textsubscript{x}
having a high kinetic inductance thanks to a low superfluid density
\citep{Peltonen2013}. Here it is virtual vortex tunneling through
the narrow line that is the conjugate of the charge on the junction
capacitance. It was however noted that formation of sub-gap states
should be avoided in spite of disorder, as such states would introduce
dissipation and thus de-coherence. We show here that nano-scale granular
superconductors provide better solution because they are close to
a Mott transition rather than to an Anderson transition, typical of
atomic disorder, which induces a massive presence of sub-gap states.

When the mean free path $l$ of a metal is reduced by disorder to
be smaller than the BCS coherence length $\xi_{0}$, the superfluid
density $n_{s}$ can be decreased substantially. For a given value
of the current, pair velocity is much increased and the kinetic energy
of the Cooper pairs becomes the dominant factor of the self-inductance
of a stripe.

In a stripe of length $L$, width $w$ and thickness $t$ the kinetic
energy of the superfluid with density $n_{s}$ and velocity $v_{s}$
is stored as inductive energy

\begin{equation}
n_{s}Lwt\frac{1}{2}(2m)v_{s}^{2}=\frac{1}{2}L_{k}I^{2}
\end{equation}
since $J_{s}=I_{s}/wt=n_{s}(2e)v_{s}$ and the penetration depth definition
is $\lambda^{2}=\frac{2m}{\mu_{0}n_{s}(2e)^{2}}$ we obtain for the
inductance
\begin{equation}
L_{k/\oblong}=\mu_{0}\frac{\lambda^{2}}{t}
\end{equation}
where $L_{k/\oblong}$ is the kinetic inductance divided by the number
of squares $L/w$. Deep in the dirty limit, defined as $l\ll\xi_{0}$,
$\lambda^{2}\simeq\lambda_{L}^{2}\xi_{0}/l$, where $\lambda_{L}$
is the London penetration depth \citep{Tinkham2004}. Here $\lambda_{L}$
is the penetration depth in the clean limit where $n_{s}=n/2$. Using
the dirty limit expression for $\lambda^{2}$ and the BCS relation
for $\xi_{0}=\hbar v_{F}/\pi\Delta$ where $v_{F}$ is the Fermi velocity
and $\Delta$ is the SC gap, and taking into account that $l=v_{F}\tau$
where $\tau$ is the relaxation time, we obtain 

\begin{equation}
L_{k/\oblong}=\frac{\hbar\rho_{n}}{\pi\Delta t}\label{eq: LkDL}
\end{equation}
where $\rho_{n}=m/ne^{2}\tau$ is the normal state resistivity. In
the BCS weak coupling $2\Delta=3.53k_{B}T_{c}$ \citep{Bardeen1957}
one can calculate $L_{k/\oblong}$ from the values of the sheet resistance
$R_{\oblong}=\rho_{n}/t$ and the critical temperature.

A high inductance can be reached by increasing the value of the sheet
resistance. This can be achieved by decreasing the film thickness
(for example Nb \citep{Annunziata2009,Annunziata2010}), and/or by
using a highly disordered metal such as NbN\textsubscript{x} \citep{Niepce2019,Peltonen2013},
Nb\textsubscript{x}Si\textsubscript{1-x} \citep{leSueur2018}, TiN
\citep{Leduc2010,Vissers2010,Swenson2013,CoumouJune} and NbTiN \citep{Niepce2019,Peltonen2013,leSueur2018,Leduc2010,Vissers2010,Swenson2013,CoumouJune,Samkharadze2016,Gruenhaupt2018,Maleeva2018}.
Because very thin films tend to be discontinuous the inductance values
one can reach in this way are rather limited, of the order of a few
pH per square \citep{Annunziata2010}. Much higher values, of the
order of 1 nH per square, have been reached by using strongly disordered
films approaching the metal-to-insulator (M/I) transition, such as
NbN\textsubscript{x} \citep{Peltonen2013}. However it was found
out that the quality factor of resonators made of such disordered
films tends to deteriorate with disorder \citep{Peltonen2013}. This
has so far limited the use of disordered superconducting for high
inductance devices, because higher losses mean increased de-coherence
effects thus making quantum computing impractical. 

Another material that has been considered recently for application
in QC is granular aluminum (grAl) \citep{Gruenhaupt2018,Maleeva2018,Winkel2019}.
grAl films can be prepared by thermally evaporating clean aluminum
pellets in oxygen environment, for more details see \citep{Moshe2019}.
Their structure consists of small grains separated by aluminum oxide
insulating barriers. The inter-grain coupling and resulting normal
state resistivity are controlled by the oxygen partial pressure and
Al deposition rate. High sheet resistance is reached by reducing inter-grain
coupling rather than by atomic scale disorder. The phase diagram of
$T_{c}$ vs resistivity $\rho_{n}$ has the well known ``dome''
shape, first rising to reach a maximum value which depends on the
temperature of the substrate during film growth (respectively 2.3
K at room temperature \citep{Levy-Bertrand2019,Deutscher1973a,Cohen1968,Dynes1984},
3 nm grain size \citep{Deutscher1973a}, and 3.2 K at 100 K \citep{Moshe2019,Bachar2015,Bachar2013,Pracht2016,Pracht2017,Deutscher1973},
2 nm grain size with a narrower distribution \citep{Deutscher1973,Lerer2014})
and thereafter decreasing as the metal-to-insulator transition is
approached.

We show in Fig. \ref{fig:TcNbNGral} the decrease of $T_{c}$ with
resistivity of grAl films prepared as described above on liquid nitrogen
cooled substrates. This decrease is slow. At a resistivity of $\sim7\times10^{4}\,\mu\Omega\,cm$
$T_{c}$ is still 40\% of its maximum value. By contrast, the $T_{c}$
of NbN\textsubscript{x} films collapses much faster. They become
useless for device applications at resistivity values of the order
of less than $1\times10^{4}\,\mu\Omega\,cm$. grAl based devices can
have kinetic inductance values one order of magnitude higher than
NbN\textsubscript{x} based devices, which represents a considerable
advantage. Furthermore, we show in Fig. \ref{fig:RT_Tunneling} that
the superconductivity transition remains sharp up to the highest resistivity.
\begin{center}
\begin{figure}
\centering{}\includegraphics[width=1\columnwidth]{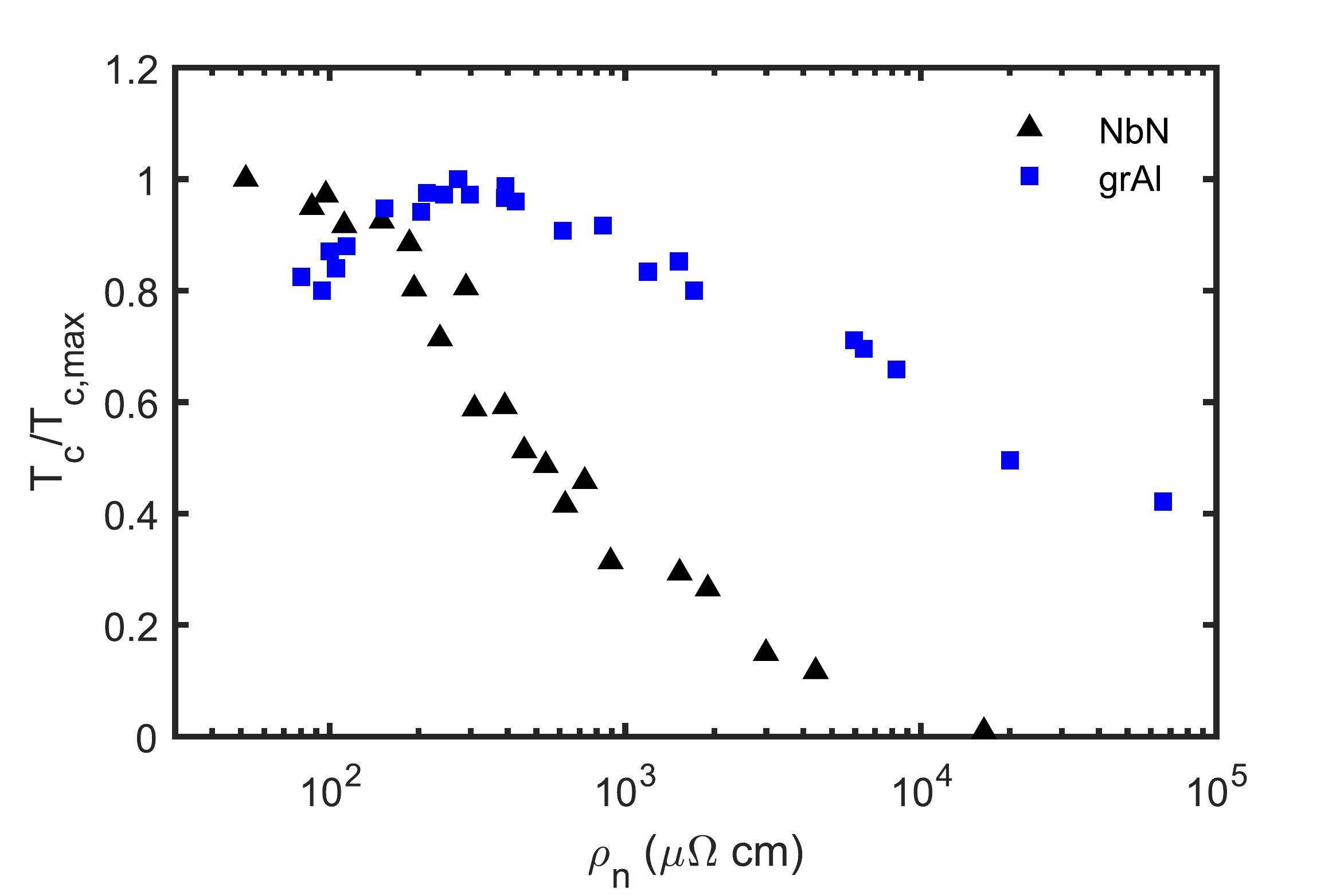}\caption{Variation of the critical temperature vs resistivity for 2 nm grain
size samples. Black triangles are NbN\protect\textsubscript{x} data
\citep{Mondal2011} and blue squares are previously published grAl
data \citep{Moshe2019} and our new data\textbf{.} The critical temperature
values are normalized to their maximum of 16.69 K and 3.25 K, respectively.
When superconductivity in NbN\protect\textsubscript{x} is quenched,
it still persists up to relatively high resistivity values in nano-scale
grAl. Note that even close to the metal-to-insulator transition the
critical temperature of grAl is above that of pure bulk aluminum,
which is marked by the value $T_{c}/T_{c,max}=0.37$ in the figure.
\label{fig:TcNbNGral}}
\end{figure}
\par\end{center}

\begin{figure}
\centering{}\includegraphics[width=1\columnwidth]{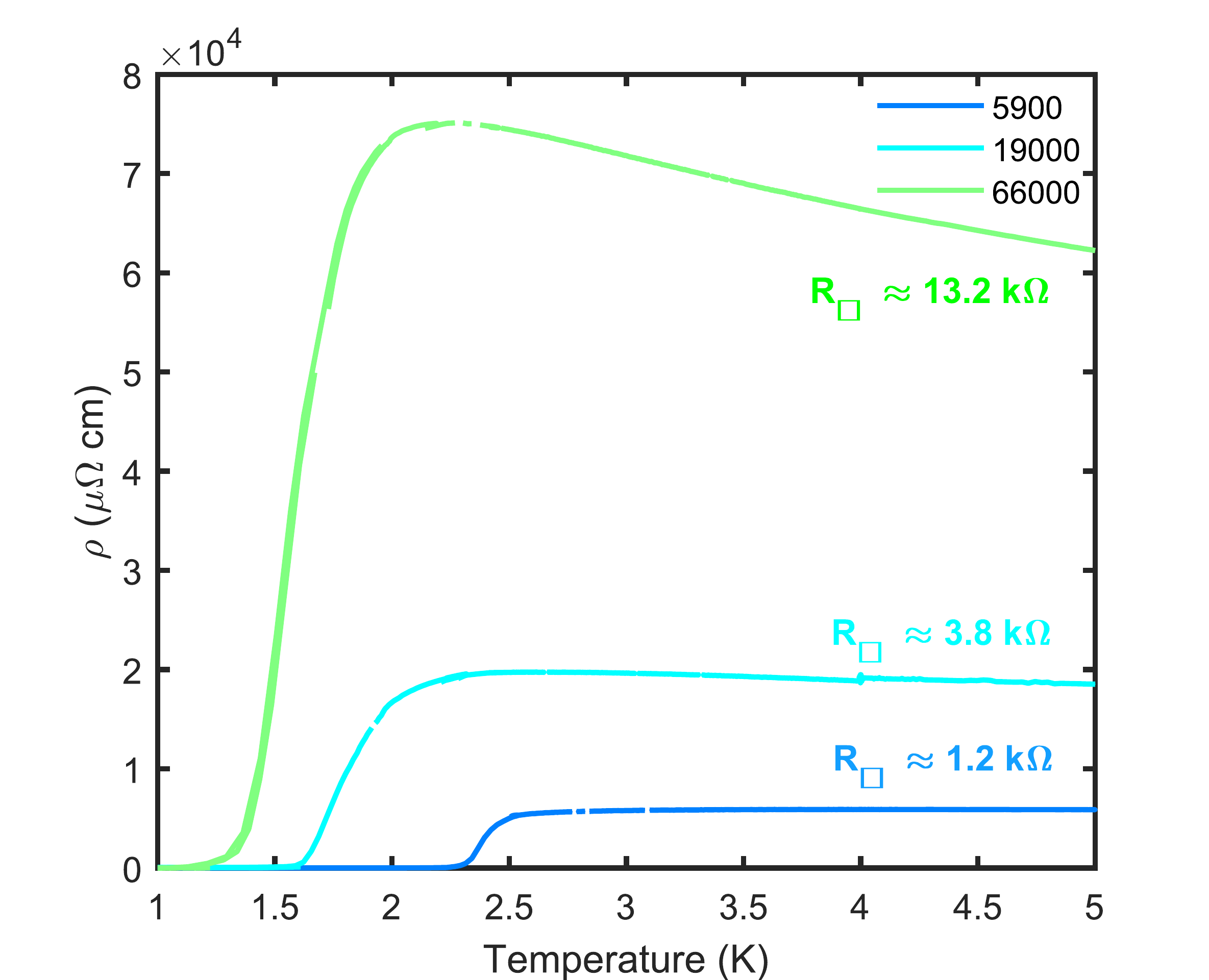}\caption{Resistance vs temperature curves of high resistivity films with 2
nm grain size. The legend corresponds to the normal state resistivity
value in $\mu\Omega\,cm$, taken as the value at 4.2 K. The corresponding
value of the normal state sheet resistance is marked near each curve,
exceeding the value $R_{\square}=h/(4e^{2})\sim6.5\,k\Omega$ for
the highest resistance sample. \label{fig:RT_Tunneling}}
\end{figure}

A further critical point concerns the level of losses, as seen for
instance in resonators. In Fig. \ref{fig:Internal-quality-factor}
we have collected values of internal quality factors reported for
a number of devices as a function of their kinetic inductance values,
with the aim of comparing devices based on atomically disordered superconductors
such as NbN\textsubscript{x} with grAl devices (here using films
deposited at room temperature \citep{Valenti2019,Gruenhaupt2018,Maleeva2018}).
The data shows a continuous decrease for the former, and no systematic
change for the later. We emphasize that the highest kinetic inductance
values reported in this figure are of about 1 $nH/\oblong$. This
is because higher values cannot be reached for NbN\textsubscript{x}
and similar atomically disorder films, for the reasons explained above.
However, in view of the high quality of our grAl films deposited at
liquid nitrogen temperature as shown Fig. \ref{fig:RT_Tunneling}
we believe that values as high as 10 $nH/\oblong$ can be reached
without increased losses.

We propose that the origin of the different behaviors of atomically
disordered and granular superconductors lies in the different nature
of their metal to insulator transition. When disorder is on the atomic
scale the transition is of the Anderson type. This is the case in
NbN\textsubscript{x} films where disorder is created by introducing
vacancies through a reduction of the nitrogen concentration \citep{Mondal2011,Chand2012}.
The density of states (DOS) of delocalized states decreases progressively
to zero as more carriers near the Fermi level become localized. If
the metal is a superconductor, the reduced DOS results in a fast decrease
of the critical temperature. At the same time the localized carriers
create sub-gap states such as two-level systems or collective modes
such as the Higgs modes \citep{Sherman2015} and increased losses
must be expected. An increased density of sub-gap states is intrinsic
to the vicinity of an Anderson transition.

The nature of the metal-to-insulator transition is quite different
in a granular metal consisting of nano-size metal crystallites weakly
coupled together. When the coupling is weak enough, the Coulomb charging
energy of the grains turns the granular system into an insulator \citep{Abeles1977,Beloborodov2007}.
By analogy with the Hubbard case \citep{Georges1996} the transition
can be of the Mott type when that energy is of the order of the effective
band width of the granular system (determined by the strength of the
inter-grain coupling), if disorder effects are not dominant. As discussed
by Beloborodov \textit{et al., }disorder will in fact dominate if
the spacing between the electronic levels in the individual grains
is small \citep{Beloborodov2007}. The spacing can be approximated
by $\delta=1/N(0)V$ \citep{Kubo1962} where $N(0)$ is the DOS at
the Fermi level and $V$ is the grain volume. That spacing is indeed
small in the Al-Ge system, in which the grain size is about 10 nm
\citep{Gerber1997}. The spacing being then of the order of 1 K. But
when the grain size is about 2 nm, the inter-level spacing is about
100 K \citep{Bachar2020}. If, in addition the grain size distribution
is narrow as is the case here, a Mott transition can be preserved.
Indeed, disorder appears to play in that case only a minor role since
superconductivity persists up to $k_{F}l$ values smaller than unity!
\citep{Moshe2019}. Additional experimental evidence for a Mott transition
in nano-scale grAl has been previously presented and discussed \citep{Bachar2015}.

\begin{table}
\centering{}%
\begin{tabular}{|c|c|c|c|c|}
\hline 
Material & $R_{\oblong}(\Omega)$ & t (nm) & $Q_{i}$ & $\begin{array}{c}
L_{k/\oblong}\\
(\text{nH}/\oblong)
\end{array}$\tabularnewline
\hline 
\hline 
grAl \citep{Valenti2019} & 20 & 20 & $1.7\times10^{5}$ & 0.016{*}\tabularnewline
\hline 
grAl \citep{Valenti2019} & 40 & 20 & $\sim3\times10^{4}$ & 0.032{*}\tabularnewline
\hline 
grAl \citep{Valenti2019} & 80 & 20 & $\sim1.5\times10^{5}$ & 0.064{*}\tabularnewline
\hline 
grAl \citep{Valenti2019} & 110 & 20 & $\sim4\times10^{5}$ & 0.088{*}\tabularnewline
\hline 
grAl \citep{Valenti2019} & 450 & 20 & $\sim10^{5}$ & 0.360{*}\tabularnewline
\hline 
grAl \citep{Valenti2019} & 800 & 20 & $\sim2\times10^{5}$ & 0.640{*}\tabularnewline
\hline 
grAl \citep{Gruenhaupt2018,Maleeva2018} & 2000 & 20 & \textasciitilde$10^{5}${} & 2\tabularnewline
\hline 
grAl \citep{Zhang2019} & 1645 & 26 & $\sim2\times10^{4}$ & 1.2\tabularnewline
\hline 
grAl \citep{Zhang2019} & 1661 & 25 & $\sim10^{4}$ & 1.2\tabularnewline
\hline 
grAl \citep{Zhang2019} & 2706 & 37 & $\sim10^{4}$ & 2\tabularnewline
\hline 
NbN\textsubscript{x} \citep{Niepce2019} & 500 & 20 & $2.5\times10^{4}$ & 0.082{*}\tabularnewline
\hline 
NbN\textsubscript{x} \citep{Peltonen2013} & 2000 & 2-3 & $\sim10^{3}$ & 1.3\tabularnewline
\hline 
Nb\textsubscript{x}Si\textsubscript{1-x} (x=0.18) \citep{leSueur2018} & 600 & 15 & $10^{3}-10^{4}$ & 0.83\tabularnewline
\hline 
TiN \citep{Leduc2010,Vissers2010} & 25 & 40 & \textasciitilde$10^{4}-10^{6}${} & 0.008{*}\tabularnewline
\hline 
TiN \citep{Swenson2013} & 45 & 22 & $8.7\times10^{5}$ & 0.031{*}\tabularnewline
\hline 
TiN \citep{CoumouJune} & 600 & 6 & \textasciitilde$10^{4}${} & 0.620\tabularnewline
\hline 
TiN \citep{Shearrow2018} & 505 & 8.9 & $(0.7-1)\times10^{5}$ & 0.234{*}\tabularnewline
\hline 
TiN \citep{Shearrow2018} & 145 & 14.2 & $(2-6)\times10^{5}$ & 0.056{*}\tabularnewline
\hline 
TiN \citep{Shearrow2018} & 21 & 49.8 & $(0.3-2)\times10^{6}$ & 0.0071{*}\tabularnewline
\hline 
TiN \citep{Shearrow2018} & 6 & 109 & $10^{5}-10^{6}$ & 0.0017{*}\tabularnewline
\hline 
NbTiN \citep{Samkharadze2016} & 250 & 8 & \textasciitilde$10^{5}${} & 0.035{*}\tabularnewline
\hline 
\end{tabular}\caption{Internal quality factors $Q_{i}$ at low circulating photon numbers,
approaching single photon regime. Asterisk marks $L_{k/\oblong}$
as estimated by Eq. \ref{eq: LkDL} with the assumption of a BCS weak
coupling ratio. Note that all grAl data shown in this table was taken
for samples deposited onto substrates held at room temperature. \label{tab:Qi}}
\end{table}

\begin{center}
\begin{figure}
\centering{}\includegraphics[width=1\columnwidth]{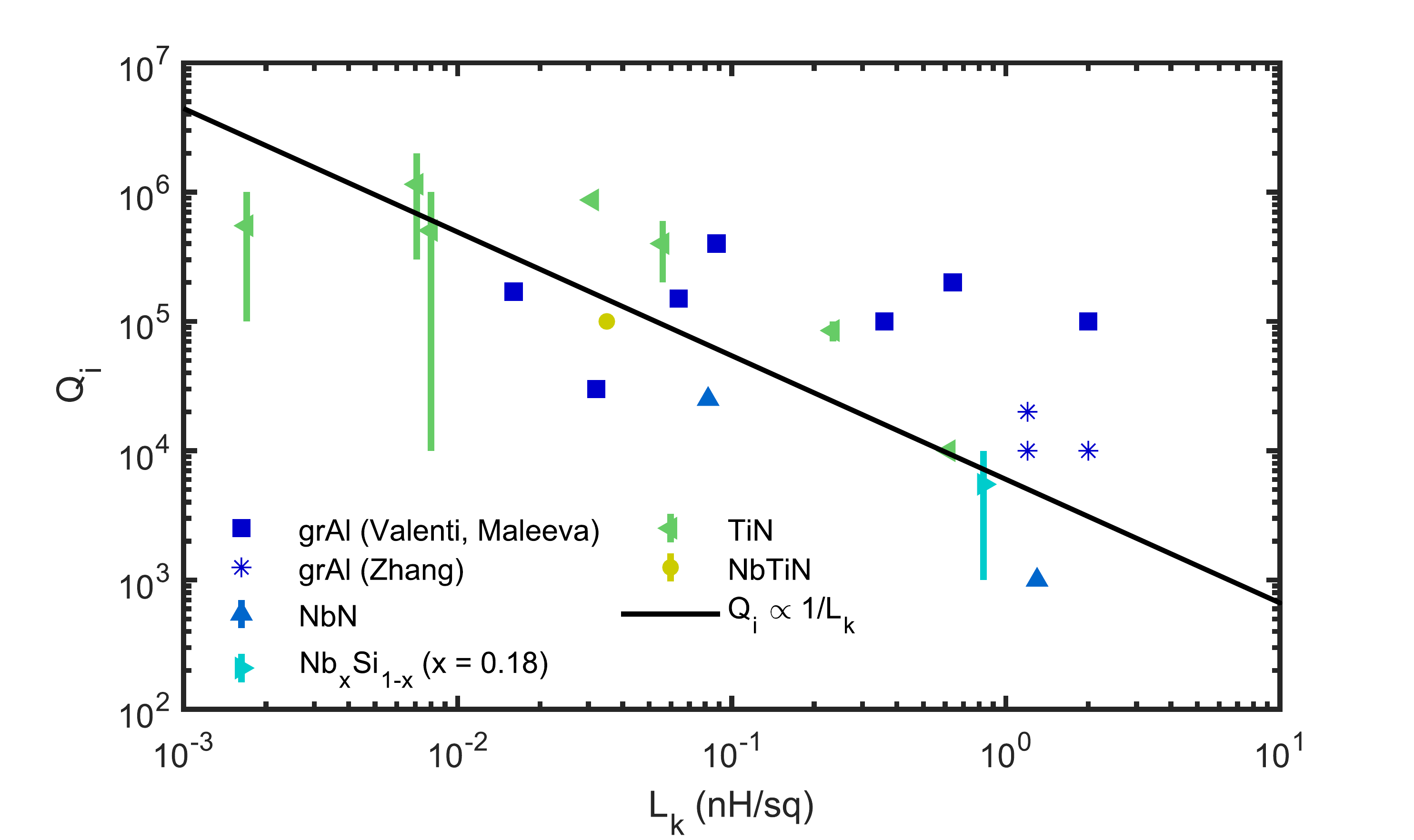}\caption{Internal quality factor $Q_{i}$ vs. sheet kinetic inductance $L_{k/\oblong}$.
For devices in which the $Q_{i}$ values are widely distributed we
represent the data as a thick vertical line. At high $L_{k/\oblong}$
values devices based on grAl films have a higher $Q_{i}$ than films
of atomically disordered superconductors. Note that the data for grAl
is from two different groups, represented by Zhang \citep{Zhang2019}
and Valenti, Maleeva \citep{Valenti2019,Maleeva2018} in the legend.
\label{fig:Internal-quality-factor}}
\end{figure}
\par\end{center}

\begin{center}
\begin{figure}
\centering{}\includegraphics[width=1\columnwidth]{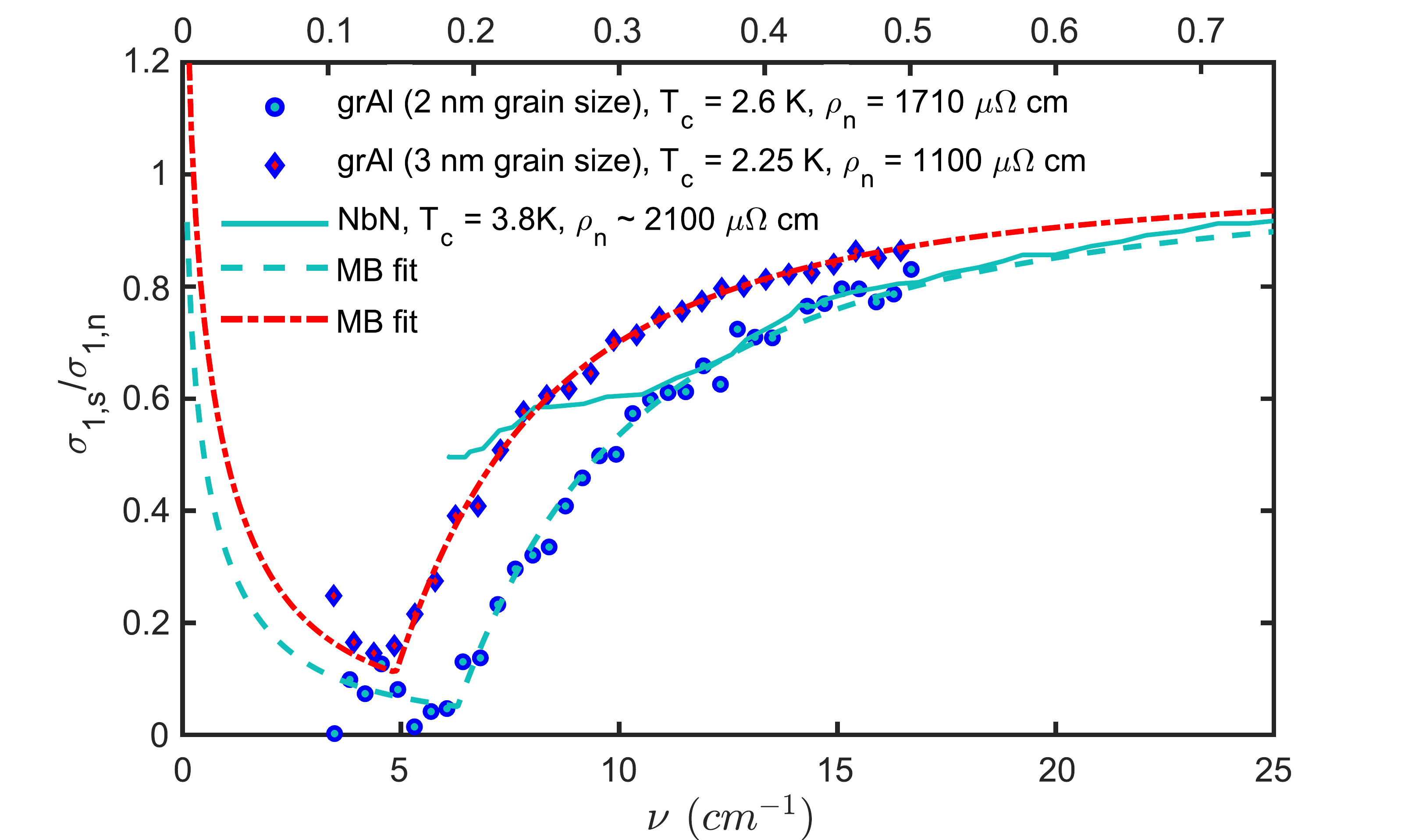}\caption{Comparison of the real part of the optical conductivity between grAl
and NbN\protect\textsubscript{x} at temperature of 1.5 K. Circles
are 2 nm grain size grAl data \citep{Moshe2019}, diamonds are 3 nm
grain size grAl data and full line is NbN\protect\textsubscript{x}
data \citep{Cheng2016}. For the grAl samples the fit to MB theory
(dashed lines) works well all the way down to the gap value, marked
by the minimum of $\sigma_{1}$ at $2\Delta$. For the NbN\protect\textsubscript{x}
sample it does not. The frequency units are given in wavenumbers which
are related to the angular frequency by $\omega=2\pi c\nu$. Top axis
frequency units are given in THz. \label{fig:sig1NbNGral}}
\end{figure}
\par\end{center}

In Fig. \ref{fig:sig1NbNGral} we compare the optical conductivity
versus frequency of the two systems for similar values of the resistivity.
In high resistivity NbN\textsubscript{x} samples the real part of
the conductivity remains high below the gap value obtained from a
fit to Mattis-Bardeen (MB) theory \citep{Mattis1958} at high frequencies,
indicating the massive presence of in-gap states. This deviation from
MB is systematic for high resistivity low $T_{c}$ NbN\textsubscript{x}
films \citep{Sherman2015,Cheng2016}. In strong contrast, in high
resistivity grAl samples the fit to MB theory remains good down to
low frequencies, for both grain sizes. The data for 3 nm grain size
sample has been taken by us with the same setup as described in \citep{Moshe2019}.
The density of sub-gap states, if present, is evidently much lower
than in NbN\textsubscript{x}. The good agreement with the MB theory
is in line with the measured temperature dependence of the change
in resonance frequency $\delta f_{0}/f_{0}$ for a series of resonators
made of grAl, with resistvities up to $1600\,\mu\Omega\,cm$, deposited
onto substrates held at room temperature \citep{Valenti2019}.

The internal quality factor can be approximated by the expression
$Q_{i}\simeq\sigma_{2}/\sigma_{1}$ which holds for thin film $(t\ll\lambda)$
and a kinetic induction fraction $\alpha=L_{k}/L_{g}$ of order unity
\citep{CoumouJune}. At frequencies below the gap,$\sigma_{2}\propto1/\omega$
and in the zero temperature limit the value of $\sigma_{1}$ is expected
to approach zero for $\hbar\omega<2\Delta$ (the data in Fig. 4 was
taken at $T/T_{c}$\textasciitilde 0.6-0.7 for grAl and \textasciitilde 0.4
for NbN\textsubscript{x}) \citep{Mattis1958}. Therefore, the presence
of sub-gap states for NbN\textsubscript{x} suggests that a lower quality factor would
be obtained than for grAl. A difference by two orders of magnitude
in the quality factors as can be seen in Table \ref{tab:Qi} between
grAl and NbN\textsubscript{x} having a similar sheet resistance of 2000 \textgreek{W}
is therefore reasonable.

On the basis of these compared optical conductivity data we expect
higher losses in devices based on NbN\textsubscript{x}, and other
atomically disordered superconductors, than in devices using nano-scale
grAl. This is borne out by the respective quality factor values of
resonators made of these two types of materials, see Table \ref{tab:Qi}.
In atomically disordered superconductors the quality factor decreases
as the sheet resistance and the related inductance increase, while
resonators made of grAl have both a high inductance and a high quality
factor.

Regarding the possibility of other Mott granular SC, we suggest granular
materials having a level spacing of the order of 100 K. The advantage
of grAl is that this criterion is established naturally during the
sample evaporation, without any need for expensive techniques. Al-Ge
films deposited at low temperature are also a possibility \citep{Fontaine1972}.

In summary, we suggest that the nature of the M/I transition is crucial
for achieving both high kinetic inductance and high quality factor
devices. Atomically disordered SC are less suitable for this purpose,
as evident from the low quality factors which we believe result from
massive presence of sub-gap states seen from optical conductivity
data. In contrast, nano-scale superconductors, such as grAl, are more
suitable for this purpose. Their higher quality factor at large kinetic
inductance values results we believe from the absence of intrinsic
sub-gap states. We have interpreted this difference as being due to
the different nature of the metal-to-insulator transition in these
two systems, being of the Anderson type in NbN\textsubscript{x}and
of the Mott type in nano-scale grAl. We suggest therefore that high
inductance devices based on superconducting granular films, consisting
of grains of only a few nano-meters with a narrow size distribution
so as to reduce disorder effects, should be further considered for
implementation in quantum circuits.

We acknowledge fruitful discussions with Ioan Pop, Florence Levy-Bertrand,
Marc Scheffler, Nimrod Bachar and Christoph Strunk, held at the Superconducting
Kinetic Inductances 2019 workshop in Bad Honnef, hosted by the Wilhelm
and Else Heraeus-Foundation.

\section*{DATA AVAILABILITY}

The data that supports the findings of this study are available within
the article.

\bibliographystyle{apsrev4-1}

\begin{thebibliography}{45}%
\makeatletter
\providecommand \@ifxundefined [1]{%
 \@ifx{#1\undefined}
}%
\providecommand \@ifnum [1]{%
 \ifnum #1\expandafter \@firstoftwo
 \else \expandafter \@secondoftwo
 \fi
}%
\providecommand \@ifx [1]{%
 \ifx #1\expandafter \@firstoftwo
 \else \expandafter \@secondoftwo
 \fi
}%
\providecommand \natexlab [1]{#1}%
\providecommand \enquote  [1]{``#1''}%
\providecommand \bibnamefont  [1]{#1}%
\providecommand \bibfnamefont [1]{#1}%
\providecommand \citenamefont [1]{#1}%
\providecommand \href@noop [0]{\@secondoftwo}%
\providecommand \href [0]{\begingroup \@sanitize@url \@href}%
\providecommand \@href[1]{\@@startlink{#1}\@@href}%
\providecommand \@@href[1]{\endgroup#1\@@endlink}%
\providecommand \@sanitize@url [0]{\catcode `\\12\catcode `\$12\catcode
  `\&12\catcode `\#12\catcode `\^12\catcode `\_12\catcode `\%12\relax}%
\providecommand \@@startlink[1]{}%
\providecommand \@@endlink[0]{}%
\providecommand \url  [0]{\begingroup\@sanitize@url \@url }%
\providecommand \@url [1]{\endgroup\@href {#1}{\urlprefix }}%
\providecommand \urlprefix  [0]{URL }%
\providecommand \Eprint [0]{\href }%
\providecommand \doibase [0]{http://dx.doi.org/}%
\providecommand \selectlanguage [0]{\@gobble}%
\providecommand \bibinfo  [0]{\@secondoftwo}%
\providecommand \bibfield  [0]{\@secondoftwo}%
\providecommand \translation [1]{[#1]}%
\providecommand \BibitemOpen [0]{}%
\providecommand \bibitemStop [0]{}%
\providecommand \bibitemNoStop [0]{.\EOS\space}%
\providecommand \EOS [0]{\spacefactor3000\relax}%
\providecommand \BibitemShut  [1]{\csname bibitem#1\endcsname}%
\let\auto@bib@innerbib\@empty
\bibitem [{\citenamefont {Bouchiat}\ \emph {et~al.}(1998)\citenamefont
  {Bouchiat}, \citenamefont {Vion}, \citenamefont {Joyez}, \citenamefont
  {Esteve},\ and\ \citenamefont {Devoret}}]{Bouchiat1998}%
  \BibitemOpen
  \bibfield  {author} {\bibinfo {author} {\bibfnamefont {V.}~\bibnamefont
  {Bouchiat}}, \bibinfo {author} {\bibfnamefont {D.}~\bibnamefont {Vion}},
  \bibinfo {author} {\bibfnamefont {P.}~\bibnamefont {Joyez}}, \bibinfo
  {author} {\bibfnamefont {D.}~\bibnamefont {Esteve}}, \ and\ \bibinfo {author}
  {\bibfnamefont {M.~H.}\ \bibnamefont {Devoret}},\ }\href {\doibase
  10.1238/physica.topical.076a00165} {\bibfield  {journal} {\bibinfo  {journal}
  {Physica Scripta}\ }\textbf {\bibinfo {volume} {T76}},\ \bibinfo {pages}
  {165} (\bibinfo {year} {1998})}\BibitemShut {NoStop}%
\bibitem [{\citenamefont {Nakamura}\ \emph {et~al.}(1999)\citenamefont
  {Nakamura}, \citenamefont {Pashkin},\ and\ \citenamefont
  {Tsai}}]{Nakamura1999}%
  \BibitemOpen
  \bibfield  {author} {\bibinfo {author} {\bibfnamefont {Y.}~\bibnamefont
  {Nakamura}}, \bibinfo {author} {\bibfnamefont {Y.~A.}\ \bibnamefont
  {Pashkin}}, \ and\ \bibinfo {author} {\bibfnamefont {J.~S.}\ \bibnamefont
  {Tsai}},\ }\href {https://doi.org/10.1038/19718} {\bibfield  {journal}
  {\bibinfo  {journal} {Nature}\ }\textbf {\bibinfo {volume} {398}},\ \bibinfo
  {pages} {786} (\bibinfo {year} {1999})}\BibitemShut {NoStop}%
\bibitem [{\citenamefont {Mooij}\ \emph {et~al.}(1999)\citenamefont {Mooij},
  \citenamefont {Orlando}, \citenamefont {Levitov}, \citenamefont {Tian},
  \citenamefont {van~der Wal},\ and\ \citenamefont {Lloyd}}]{Mooij1999}%
  \BibitemOpen
  \bibfield  {author} {\bibinfo {author} {\bibfnamefont {J.~E.}\ \bibnamefont
  {Mooij}}, \bibinfo {author} {\bibfnamefont {T.~P.}\ \bibnamefont {Orlando}},
  \bibinfo {author} {\bibfnamefont {L.}~\bibnamefont {Levitov}}, \bibinfo
  {author} {\bibfnamefont {L.}~\bibnamefont {Tian}}, \bibinfo {author}
  {\bibfnamefont {C.~H.}\ \bibnamefont {van~der Wal}}, \ and\ \bibinfo {author}
  {\bibfnamefont {S.}~\bibnamefont {Lloyd}},\ }\href {\doibase
  10.1126/science.285.5430.1036} {\bibfield  {journal} {\bibinfo  {journal}
  {Science}\ }\textbf {\bibinfo {volume} {285}},\ \bibinfo {pages} {1036}
  (\bibinfo {year} {1999})},\ \Eprint
  {http://arxiv.org/abs/https://science.sciencemag.org/content/285/5430/1036.full.pdf}
  {https://science.sciencemag.org/content/285/5430/1036.full.pdf} \BibitemShut
  {NoStop}%
\bibitem [{\citenamefont {Manucharyan}\ \emph {et~al.}(2009)\citenamefont
  {Manucharyan}, \citenamefont {Koch}, \citenamefont {Glazman},\ and\
  \citenamefont {Devoret}}]{Manucharyan2009}%
  \BibitemOpen
  \bibfield  {author} {\bibinfo {author} {\bibfnamefont {V.~E.}\ \bibnamefont
  {Manucharyan}}, \bibinfo {author} {\bibfnamefont {J.}~\bibnamefont {Koch}},
  \bibinfo {author} {\bibfnamefont {L.~I.}\ \bibnamefont {Glazman}}, \ and\
  \bibinfo {author} {\bibfnamefont {M.~H.}\ \bibnamefont {Devoret}},\ }\href
  {\doibase 10.1126/science.1175552} {\bibfield  {journal} {\bibinfo  {journal}
  {Science}\ }\textbf {\bibinfo {volume} {326}},\ \bibinfo {pages} {113}
  (\bibinfo {year} {2009})},\ \Eprint
  {http://arxiv.org/abs/https://science.sciencemag.org/content/326/5949/113.full.pdf}
  {https://science.sciencemag.org/content/326/5949/113.full.pdf} \BibitemShut
  {NoStop}%
\bibitem [{\citenamefont {Peltonen}\ \emph {et~al.}(2013)\citenamefont
  {Peltonen}, \citenamefont {Astafiev}, \citenamefont {Korneeva}, \citenamefont
  {Voronov}, \citenamefont {Korneev}, \citenamefont {Charaev}, \citenamefont
  {Semenov}, \citenamefont {Golt'sman}, \citenamefont {Ioffe}, \citenamefont
  {Klapwijk},\ and\ \citenamefont {Tsai}}]{Peltonen2013}%
  \BibitemOpen
  \bibfield  {author} {\bibinfo {author} {\bibfnamefont {J.~T.}\ \bibnamefont
  {Peltonen}}, \bibinfo {author} {\bibfnamefont {O.~V.}\ \bibnamefont
  {Astafiev}}, \bibinfo {author} {\bibfnamefont {Y.~P.}\ \bibnamefont
  {Korneeva}}, \bibinfo {author} {\bibfnamefont {B.~M.}\ \bibnamefont
  {Voronov}}, \bibinfo {author} {\bibfnamefont {A.~A.}\ \bibnamefont
  {Korneev}}, \bibinfo {author} {\bibfnamefont {I.~M.}\ \bibnamefont
  {Charaev}}, \bibinfo {author} {\bibfnamefont {A.~V.}\ \bibnamefont
  {Semenov}}, \bibinfo {author} {\bibfnamefont {G.~N.}\ \bibnamefont
  {Golt'sman}}, \bibinfo {author} {\bibfnamefont {L.~B.}\ \bibnamefont
  {Ioffe}}, \bibinfo {author} {\bibfnamefont {T.~M.}\ \bibnamefont {Klapwijk}},
  \ and\ \bibinfo {author} {\bibfnamefont {J.~S.}\ \bibnamefont {Tsai}},\
  }\href {\doibase 10.1103/PhysRevB.88.220506} {\bibfield  {journal} {\bibinfo
  {journal} {Phys. Rev. B}\ }\textbf {\bibinfo {volume} {88}},\ \bibinfo
  {pages} {220506} (\bibinfo {year} {2013})}\BibitemShut {NoStop}%
\bibitem [{\citenamefont {Tinkham}(2004)}]{Tinkham2004}%
  \BibitemOpen
  \bibfield  {author} {\bibinfo {author} {\bibfnamefont {M.}~\bibnamefont
  {Tinkham}},\ }\href {https://books.google.co.il/books?id=k6AO9nRYbioC} {\emph
  {\bibinfo {title} {Introduction to Superconductivity: Second Edition}}},\
  Dover Books on Physics\ (\bibinfo  {publisher} {Dover Publications},\
  \bibinfo {year} {2004})\BibitemShut {NoStop}%
\bibitem [{\citenamefont {Bardeen}\ \emph {et~al.}(1957)\citenamefont
  {Bardeen}, \citenamefont {Cooper},\ and\ \citenamefont
  {Schrieffer}}]{Bardeen1957}%
  \BibitemOpen
  \bibfield  {author} {\bibinfo {author} {\bibfnamefont {J.}~\bibnamefont
  {Bardeen}}, \bibinfo {author} {\bibfnamefont {L.~N.}\ \bibnamefont {Cooper}},
  \ and\ \bibinfo {author} {\bibfnamefont {J.~R.}\ \bibnamefont {Schrieffer}},\
  }\href {\doibase 10.1103/PhysRev.108.1175} {\bibfield  {journal} {\bibinfo
  {journal} {Physical Review}\ }\textbf {\bibinfo {volume} {108}},\ \bibinfo
  {pages} {1175} (\bibinfo {year} {1957})}\BibitemShut {NoStop}%
\bibitem [{\citenamefont {{Annunziata}}\ \emph {et~al.}(2009)\citenamefont
  {{Annunziata}}, \citenamefont {{Santavicca}}, \citenamefont {{Chudow}},
  \citenamefont {{Frunzio}}, \citenamefont {{Rooks}}, \citenamefont
  {{Frydman}},\ and\ \citenamefont {{Prober}}}]{Annunziata2009}%
  \BibitemOpen
  \bibfield  {author} {\bibinfo {author} {\bibfnamefont {A.~J.}\ \bibnamefont
  {{Annunziata}}}, \bibinfo {author} {\bibfnamefont {D.~F.}\ \bibnamefont
  {{Santavicca}}}, \bibinfo {author} {\bibfnamefont {J.~D.}\ \bibnamefont
  {{Chudow}}}, \bibinfo {author} {\bibfnamefont {L.}~\bibnamefont {{Frunzio}}},
  \bibinfo {author} {\bibfnamefont {M.~J.}\ \bibnamefont {{Rooks}}}, \bibinfo
  {author} {\bibfnamefont {A.}~\bibnamefont {{Frydman}}}, \ and\ \bibinfo
  {author} {\bibfnamefont {D.~E.}\ \bibnamefont {{Prober}}},\ }\href {\doibase
  10.1109/TASC.2009.2018740} {\bibfield  {journal} {\bibinfo  {journal} {IEEE
  Transactions on Applied Superconductivity}\ }\textbf {\bibinfo {volume}
  {19}},\ \bibinfo {pages} {327} (\bibinfo {year} {2009})}\BibitemShut
  {NoStop}%
\bibitem [{\citenamefont {Annunziata}\ \emph {et~al.}(2010)\citenamefont
  {Annunziata}, \citenamefont {Santavicca}, \citenamefont {Frunzio},
  \citenamefont {Catelani}, \citenamefont {Rooks}, \citenamefont {Frydman},\
  and\ \citenamefont {Prober}}]{Annunziata2010}%
  \BibitemOpen
  \bibfield  {author} {\bibinfo {author} {\bibfnamefont {A.~J.}\ \bibnamefont
  {Annunziata}}, \bibinfo {author} {\bibfnamefont {D.~F.}\ \bibnamefont
  {Santavicca}}, \bibinfo {author} {\bibfnamefont {L.}~\bibnamefont {Frunzio}},
  \bibinfo {author} {\bibfnamefont {G.}~\bibnamefont {Catelani}}, \bibinfo
  {author} {\bibfnamefont {M.~J.}\ \bibnamefont {Rooks}}, \bibinfo {author}
  {\bibfnamefont {A.}~\bibnamefont {Frydman}}, \ and\ \bibinfo {author}
  {\bibfnamefont {D.~E.}\ \bibnamefont {Prober}},\ }\href {\doibase
  10.1088/0957-4484/21/44/445202} {\bibfield  {journal} {\bibinfo  {journal}
  {Nanotechnology}\ }\textbf {\bibinfo {volume} {21}},\ \bibinfo {pages}
  {445202} (\bibinfo {year} {2010})}\BibitemShut {NoStop}%
\bibitem [{\citenamefont {Niepce}\ \emph {et~al.}(2019)\citenamefont {Niepce},
  \citenamefont {Burnett},\ and\ \citenamefont {Bylander}}]{Niepce2019}%
  \BibitemOpen
  \bibfield  {author} {\bibinfo {author} {\bibfnamefont {D.}~\bibnamefont
  {Niepce}}, \bibinfo {author} {\bibfnamefont {J.}~\bibnamefont {Burnett}}, \
  and\ \bibinfo {author} {\bibfnamefont {J.}~\bibnamefont {Bylander}},\ }\href
  {\doibase 10.1103/PhysRevApplied.11.044014} {\bibfield  {journal} {\bibinfo
  {journal} {Phys. Rev. Applied}\ }\textbf {\bibinfo {volume} {11}},\ \bibinfo
  {pages} {044014} (\bibinfo {year} {2019})}\BibitemShut {NoStop}%
\bibitem [{\citenamefont {{le Sueur}}\ \emph {et~al.}(2018)\citenamefont {{le
  Sueur}}, \citenamefont {{Svilans}}, \citenamefont {{Bourlet}}, \citenamefont
  {{Murani}}, \citenamefont {{Berg{\'e}}}, \citenamefont {{Dumoulin}},\ and\
  \citenamefont {{Joyez}}}]{leSueur2018}%
  \BibitemOpen
  \bibfield  {author} {\bibinfo {author} {\bibfnamefont {H.}~\bibnamefont {{le
  Sueur}}}, \bibinfo {author} {\bibfnamefont {A.}~\bibnamefont {{Svilans}}},
  \bibinfo {author} {\bibfnamefont {N.}~\bibnamefont {{Bourlet}}}, \bibinfo
  {author} {\bibfnamefont {A.}~\bibnamefont {{Murani}}}, \bibinfo {author}
  {\bibfnamefont {L.}~\bibnamefont {{Berg{\'e}}}}, \bibinfo {author}
  {\bibfnamefont {L.}~\bibnamefont {{Dumoulin}}}, \ and\ \bibinfo {author}
  {\bibfnamefont {P.}~\bibnamefont {{Joyez}}},\ }\href@noop {} {\bibfield
  {journal} {\bibinfo  {journal} {arXiv e-prints}\ ,\ \bibinfo {eid}
  {arXiv:1810.12801}} (\bibinfo {year} {2018})},\ \Eprint
  {http://arxiv.org/abs/1810.12801} {arXiv:1810.12801 [cond-mat.supr-con]}
  \BibitemShut {NoStop}%
\bibitem [{\citenamefont {Leduc}\ \emph {et~al.}(2010)\citenamefont {Leduc},
  \citenamefont {Bumble}, \citenamefont {Day}, \citenamefont {Eom},
  \citenamefont {Gao}, \citenamefont {Golwala}, \citenamefont {Mazin},
  \citenamefont {McHugh}, \citenamefont {Merrill}, \citenamefont {Moore},
  \citenamefont {Noroozian}, \citenamefont {Turner},\ and\ \citenamefont
  {Zmuidzinas}}]{Leduc2010}%
  \BibitemOpen
  \bibfield  {author} {\bibinfo {author} {\bibfnamefont {H.~G.}\ \bibnamefont
  {Leduc}}, \bibinfo {author} {\bibfnamefont {B.}~\bibnamefont {Bumble}},
  \bibinfo {author} {\bibfnamefont {P.~K.}\ \bibnamefont {Day}}, \bibinfo
  {author} {\bibfnamefont {B.~H.}\ \bibnamefont {Eom}}, \bibinfo {author}
  {\bibfnamefont {J.}~\bibnamefont {Gao}}, \bibinfo {author} {\bibfnamefont
  {S.}~\bibnamefont {Golwala}}, \bibinfo {author} {\bibfnamefont {B.~A.}\
  \bibnamefont {Mazin}}, \bibinfo {author} {\bibfnamefont {S.}~\bibnamefont
  {McHugh}}, \bibinfo {author} {\bibfnamefont {A.}~\bibnamefont {Merrill}},
  \bibinfo {author} {\bibfnamefont {D.~C.}\ \bibnamefont {Moore}}, \bibinfo
  {author} {\bibfnamefont {O.}~\bibnamefont {Noroozian}}, \bibinfo {author}
  {\bibfnamefont {A.~D.}\ \bibnamefont {Turner}}, \ and\ \bibinfo {author}
  {\bibfnamefont {J.}~\bibnamefont {Zmuidzinas}},\ }\href {\doibase
  10.1063/1.3480420} {\bibfield  {journal} {\bibinfo  {journal} {Applied
  Physics Letters}\ }\textbf {\bibinfo {volume} {97}},\ \bibinfo {pages}
  {102509} (\bibinfo {year} {2010})},\ \Eprint
  {http://arxiv.org/abs/https://doi.org/10.1063/1.3480420}
  {https://doi.org/10.1063/1.3480420} \BibitemShut {NoStop}%
\bibitem [{\citenamefont {Vissers}\ \emph {et~al.}(2010)\citenamefont
  {Vissers}, \citenamefont {Gao}, \citenamefont {Wisbey}, \citenamefont {Hite},
  \citenamefont {Tsuei}, \citenamefont {Corcoles}, \citenamefont {Steffen},\
  and\ \citenamefont {Pappas}}]{Vissers2010}%
  \BibitemOpen
  \bibfield  {author} {\bibinfo {author} {\bibfnamefont {M.~R.}\ \bibnamefont
  {Vissers}}, \bibinfo {author} {\bibfnamefont {J.}~\bibnamefont {Gao}},
  \bibinfo {author} {\bibfnamefont {D.~S.}\ \bibnamefont {Wisbey}}, \bibinfo
  {author} {\bibfnamefont {D.~A.}\ \bibnamefont {Hite}}, \bibinfo {author}
  {\bibfnamefont {C.~C.}\ \bibnamefont {Tsuei}}, \bibinfo {author}
  {\bibfnamefont {A.~D.}\ \bibnamefont {Corcoles}}, \bibinfo {author}
  {\bibfnamefont {M.}~\bibnamefont {Steffen}}, \ and\ \bibinfo {author}
  {\bibfnamefont {D.~P.}\ \bibnamefont {Pappas}},\ }\href {\doibase
  10.1063/1.3517252} {\bibfield  {journal} {\bibinfo  {journal} {Applied
  Physics Letters}\ }\textbf {\bibinfo {volume} {97}},\ \bibinfo {pages}
  {232509} (\bibinfo {year} {2010})},\ \Eprint
  {http://arxiv.org/abs/https://doi.org/10.1063/1.3517252}
  {https://doi.org/10.1063/1.3517252} \BibitemShut {NoStop}%
\bibitem [{\citenamefont {Swenson}\ \emph {et~al.}(2013)\citenamefont
  {Swenson}, \citenamefont {Day}, \citenamefont {Eom}, \citenamefont {Leduc},
  \citenamefont {Llombart}, \citenamefont {McKenney}, \citenamefont
  {Noroozian},\ and\ \citenamefont {Zmuidzinas}}]{Swenson2013}%
  \BibitemOpen
  \bibfield  {author} {\bibinfo {author} {\bibfnamefont {L.~J.}\ \bibnamefont
  {Swenson}}, \bibinfo {author} {\bibfnamefont {P.~K.}\ \bibnamefont {Day}},
  \bibinfo {author} {\bibfnamefont {B.~H.}\ \bibnamefont {Eom}}, \bibinfo
  {author} {\bibfnamefont {H.~G.}\ \bibnamefont {Leduc}}, \bibinfo {author}
  {\bibfnamefont {N.}~\bibnamefont {Llombart}}, \bibinfo {author}
  {\bibfnamefont {C.~M.}\ \bibnamefont {McKenney}}, \bibinfo {author}
  {\bibfnamefont {O.}~\bibnamefont {Noroozian}}, \ and\ \bibinfo {author}
  {\bibfnamefont {J.}~\bibnamefont {Zmuidzinas}},\ }\href {\doibase
  10.1063/1.4794808} {\bibfield  {journal} {\bibinfo  {journal} {Journal of
  Applied Physics}\ }\textbf {\bibinfo {volume} {113}},\ \bibinfo {pages}
  {104501} (\bibinfo {year} {2013})},\ \Eprint
  {http://arxiv.org/abs/https://doi.org/10.1063/1.4794808}
  {https://doi.org/10.1063/1.4794808} \BibitemShut {NoStop}%
\bibitem [{\citenamefont {Coumou}\ \emph {et~al.}(2013)\citenamefont {Coumou},
  \citenamefont {Zuiddam}, \citenamefont {Driessen}, \citenamefont {de~Visser},
  \citenamefont {Baselmans},\ and\ \citenamefont {Klapwijk}}]{CoumouJune}%
  \BibitemOpen
  \bibfield  {author} {\bibinfo {author} {\bibfnamefont {P.~C. J.~J.}\
  \bibnamefont {Coumou}}, \bibinfo {author} {\bibfnamefont {M.~R.}\
  \bibnamefont {Zuiddam}}, \bibinfo {author} {\bibfnamefont {E.~F.~C.}\
  \bibnamefont {Driessen}}, \bibinfo {author} {\bibfnamefont {P.~J.}\
  \bibnamefont {de~Visser}}, \bibinfo {author} {\bibfnamefont {J.~J.~A.}\
  \bibnamefont {Baselmans}}, \ and\ \bibinfo {author} {\bibfnamefont {T.~M.}\
  \bibnamefont {Klapwijk}},\ }\href@noop {} {\bibfield  {journal} {\bibinfo
  {journal} {IEEE Transactions on Applied Superconductivity}\ }\textbf
  {\bibinfo {volume} {23}},\ \bibinfo {pages} {7500404} (\bibinfo {year}
  {2013})}\BibitemShut {NoStop}%
\bibitem [{\citenamefont {Samkharadze}\ \emph {et~al.}(2016)\citenamefont
  {Samkharadze}, \citenamefont {Bruno}, \citenamefont {Scarlino}, \citenamefont
  {Zheng}, \citenamefont {DiVincenzo}, \citenamefont {DiCarlo},\ and\
  \citenamefont {Vandersypen}}]{Samkharadze2016}%
  \BibitemOpen
  \bibfield  {author} {\bibinfo {author} {\bibfnamefont {N.}~\bibnamefont
  {Samkharadze}}, \bibinfo {author} {\bibfnamefont {A.}~\bibnamefont {Bruno}},
  \bibinfo {author} {\bibfnamefont {P.}~\bibnamefont {Scarlino}}, \bibinfo
  {author} {\bibfnamefont {G.}~\bibnamefont {Zheng}}, \bibinfo {author}
  {\bibfnamefont {D.~P.}\ \bibnamefont {DiVincenzo}}, \bibinfo {author}
  {\bibfnamefont {L.}~\bibnamefont {DiCarlo}}, \ and\ \bibinfo {author}
  {\bibfnamefont {L.~M.~K.}\ \bibnamefont {Vandersypen}},\ }\href {\doibase
  10.1103/PhysRevApplied.5.044004} {\bibfield  {journal} {\bibinfo  {journal}
  {Phys. Rev. Applied}\ }\textbf {\bibinfo {volume} {5}},\ \bibinfo {pages}
  {044004} (\bibinfo {year} {2016})}\BibitemShut {NoStop}%
\bibitem [{\citenamefont {Gr\"unhaupt}\ \emph {et~al.}(2018)\citenamefont
  {Gr\"unhaupt}, \citenamefont {Maleeva}, \citenamefont {Skacel}, \citenamefont
  {Calvo}, \citenamefont {Levy-Bertrand}, \citenamefont {Ustinov},
  \citenamefont {Rotzinger}, \citenamefont {Monfardini}, \citenamefont
  {Catelani},\ and\ \citenamefont {Pop}}]{Gruenhaupt2018}%
  \BibitemOpen
  \bibfield  {author} {\bibinfo {author} {\bibfnamefont {L.}~\bibnamefont
  {Gr\"unhaupt}}, \bibinfo {author} {\bibfnamefont {N.}~\bibnamefont
  {Maleeva}}, \bibinfo {author} {\bibfnamefont {S.~T.}\ \bibnamefont {Skacel}},
  \bibinfo {author} {\bibfnamefont {M.}~\bibnamefont {Calvo}}, \bibinfo
  {author} {\bibfnamefont {F.}~\bibnamefont {Levy-Bertrand}}, \bibinfo {author}
  {\bibfnamefont {A.~V.}\ \bibnamefont {Ustinov}}, \bibinfo {author}
  {\bibfnamefont {H.}~\bibnamefont {Rotzinger}}, \bibinfo {author}
  {\bibfnamefont {A.}~\bibnamefont {Monfardini}}, \bibinfo {author}
  {\bibfnamefont {G.}~\bibnamefont {Catelani}}, \ and\ \bibinfo {author}
  {\bibfnamefont {I.~M.}\ \bibnamefont {Pop}},\ }\href {\doibase
  10.1103/PhysRevLett.121.117001} {\bibfield  {journal} {\bibinfo  {journal}
  {Phys. Rev. Lett.}\ }\textbf {\bibinfo {volume} {121}},\ \bibinfo {pages}
  {117001} (\bibinfo {year} {2018})}\BibitemShut {NoStop}%
\bibitem [{\citenamefont {Maleeva}\ \emph {et~al.}(2018)\citenamefont
  {Maleeva}, \citenamefont {Grünhaupt}, \citenamefont {Klein}, \citenamefont
  {Levy-Bertrand}, \citenamefont {Dupre}, \citenamefont {Calvo}, \citenamefont
  {Valenti}, \citenamefont {Winkel}, \citenamefont {Friedrich}, \citenamefont
  {Wernsdorfer}, \citenamefont {Ustinov}, \citenamefont {Rotzinger},
  \citenamefont {Monfardini}, \citenamefont {Fistul},\ and\ \citenamefont
  {Pop}}]{Maleeva2018}%
  \BibitemOpen
  \bibfield  {author} {\bibinfo {author} {\bibfnamefont {N.}~\bibnamefont
  {Maleeva}}, \bibinfo {author} {\bibfnamefont {L.}~\bibnamefont {Grünhaupt}},
  \bibinfo {author} {\bibfnamefont {T.}~\bibnamefont {Klein}}, \bibinfo
  {author} {\bibfnamefont {F.}~\bibnamefont {Levy-Bertrand}}, \bibinfo {author}
  {\bibfnamefont {O.}~\bibnamefont {Dupre}}, \bibinfo {author} {\bibfnamefont
  {M.}~\bibnamefont {Calvo}}, \bibinfo {author} {\bibfnamefont
  {F.}~\bibnamefont {Valenti}}, \bibinfo {author} {\bibfnamefont
  {P.}~\bibnamefont {Winkel}}, \bibinfo {author} {\bibfnamefont
  {F.}~\bibnamefont {Friedrich}}, \bibinfo {author} {\bibfnamefont
  {W.}~\bibnamefont {Wernsdorfer}}, \bibinfo {author} {\bibfnamefont {A.~V.}\
  \bibnamefont {Ustinov}}, \bibinfo {author} {\bibfnamefont {H.}~\bibnamefont
  {Rotzinger}}, \bibinfo {author} {\bibfnamefont {A.}~\bibnamefont
  {Monfardini}}, \bibinfo {author} {\bibfnamefont {M.~V.}\ \bibnamefont
  {Fistul}}, \ and\ \bibinfo {author} {\bibfnamefont {I.~M.}\ \bibnamefont
  {Pop}},\ }\href {https://doi.org/10.1038/s41467-018-06386-9} {\bibfield
  {journal} {\bibinfo  {journal} {Nature Communications}\ }\textbf {\bibinfo
  {volume} {9}},\ \bibinfo {pages} {3889} (\bibinfo {year} {2018})}\BibitemShut
  {NoStop}%
\bibitem [{\citenamefont {{Winkel}}\ \emph {et~al.}(2019)\citenamefont
  {{Winkel}}, \citenamefont {{Borisov}}, \citenamefont {{Gr{\"u}nhaupt}},
  \citenamefont {{Rieger}}, \citenamefont {{Spiecker}}, \citenamefont
  {{Valenti}}, \citenamefont {{Ustinov}}, \citenamefont {{Wernsdorfer}},\ and\
  \citenamefont {{Pop}}}]{Winkel2019}%
  \BibitemOpen
  \bibfield  {author} {\bibinfo {author} {\bibfnamefont {P.}~\bibnamefont
  {{Winkel}}}, \bibinfo {author} {\bibfnamefont {K.}~\bibnamefont {{Borisov}}},
  \bibinfo {author} {\bibfnamefont {L.}~\bibnamefont {{Gr{\"u}nhaupt}}},
  \bibinfo {author} {\bibfnamefont {D.}~\bibnamefont {{Rieger}}}, \bibinfo
  {author} {\bibfnamefont {M.}~\bibnamefont {{Spiecker}}}, \bibinfo {author}
  {\bibfnamefont {F.}~\bibnamefont {{Valenti}}}, \bibinfo {author}
  {\bibfnamefont {A.~V.}\ \bibnamefont {{Ustinov}}}, \bibinfo {author}
  {\bibfnamefont {W.}~\bibnamefont {{Wernsdorfer}}}, \ and\ \bibinfo {author}
  {\bibfnamefont {I.~M.}\ \bibnamefont {{Pop}}},\ }\href@noop {} {\bibfield
  {journal} {\bibinfo  {journal} {arXiv e-prints}\ ,\ \bibinfo {eid}
  {arXiv:1911.02333}} (\bibinfo {year} {2019})},\ \Eprint
  {http://arxiv.org/abs/1911.02333} {arXiv:1911.02333 [quant-ph]} \BibitemShut
  {NoStop}%
\bibitem [{\citenamefont {Moshe}\ \emph {et~al.}(2019)\citenamefont {Moshe},
  \citenamefont {Farber},\ and\ \citenamefont {Deutscher}}]{Moshe2019}%
  \BibitemOpen
  \bibfield  {author} {\bibinfo {author} {\bibfnamefont {A.~G.}\ \bibnamefont
  {Moshe}}, \bibinfo {author} {\bibfnamefont {E.}~\bibnamefont {Farber}}, \
  and\ \bibinfo {author} {\bibfnamefont {G.}~\bibnamefont {Deutscher}},\ }\href
  {\doibase 10.1103/PhysRevB.99.224503} {\bibfield  {journal} {\bibinfo
  {journal} {Phys. Rev. B}\ }\textbf {\bibinfo {volume} {99}},\ \bibinfo
  {pages} {224503} (\bibinfo {year} {2019})}\BibitemShut {NoStop}%
\bibitem [{\citenamefont {Levy-Bertrand}\ \emph {et~al.}(2019)\citenamefont
  {Levy-Bertrand}, \citenamefont {Klein}, \citenamefont {Grenet}, \citenamefont
  {Dupr\'e}, \citenamefont {Beno\^{\i}t}, \citenamefont {Bideaud},
  \citenamefont {Bourrion}, \citenamefont {Calvo}, \citenamefont {Catalano},
  \citenamefont {Gomez}, \citenamefont {Goupy}, \citenamefont {Gr\"unhaupt},
  \citenamefont {Luepke}, \citenamefont {Maleeva}, \citenamefont {Valenti},
  \citenamefont {Pop},\ and\ \citenamefont {Monfardini}}]{Levy-Bertrand2019}%
  \BibitemOpen
  \bibfield  {author} {\bibinfo {author} {\bibfnamefont {F.}~\bibnamefont
  {Levy-Bertrand}}, \bibinfo {author} {\bibfnamefont {T.}~\bibnamefont
  {Klein}}, \bibinfo {author} {\bibfnamefont {T.}~\bibnamefont {Grenet}},
  \bibinfo {author} {\bibfnamefont {O.}~\bibnamefont {Dupr\'e}}, \bibinfo
  {author} {\bibfnamefont {A.}~\bibnamefont {Beno\^{\i}t}}, \bibinfo {author}
  {\bibfnamefont {A.}~\bibnamefont {Bideaud}}, \bibinfo {author} {\bibfnamefont
  {O.}~\bibnamefont {Bourrion}}, \bibinfo {author} {\bibfnamefont
  {M.}~\bibnamefont {Calvo}}, \bibinfo {author} {\bibfnamefont
  {A.}~\bibnamefont {Catalano}}, \bibinfo {author} {\bibfnamefont
  {A.}~\bibnamefont {Gomez}}, \bibinfo {author} {\bibfnamefont
  {J.}~\bibnamefont {Goupy}}, \bibinfo {author} {\bibfnamefont
  {L.}~\bibnamefont {Gr\"unhaupt}}, \bibinfo {author} {\bibfnamefont {U.~v.}\
  \bibnamefont {Luepke}}, \bibinfo {author} {\bibfnamefont {N.}~\bibnamefont
  {Maleeva}}, \bibinfo {author} {\bibfnamefont {F.}~\bibnamefont {Valenti}},
  \bibinfo {author} {\bibfnamefont {I.~M.}\ \bibnamefont {Pop}}, \ and\
  \bibinfo {author} {\bibfnamefont {A.}~\bibnamefont {Monfardini}},\ }\href
  {\doibase 10.1103/PhysRevB.99.094506} {\bibfield  {journal} {\bibinfo
  {journal} {Phys. Rev. B}\ }\textbf {\bibinfo {volume} {99}},\ \bibinfo
  {pages} {094506} (\bibinfo {year} {2019})}\BibitemShut {NoStop}%
\bibitem [{\citenamefont {Deutscher}\ \emph
  {et~al.}(1973{\natexlab{a}})\citenamefont {Deutscher}, \citenamefont
  {Fenichel}, \citenamefont {Gershenson}, \citenamefont {Grunbaum},\ and\
  \citenamefont {Ovadyahu}}]{Deutscher1973a}%
  \BibitemOpen
  \bibfield  {author} {\bibinfo {author} {\bibfnamefont {G.}~\bibnamefont
  {Deutscher}}, \bibinfo {author} {\bibfnamefont {H.}~\bibnamefont {Fenichel}},
  \bibinfo {author} {\bibfnamefont {M.}~\bibnamefont {Gershenson}}, \bibinfo
  {author} {\bibfnamefont {E.}~\bibnamefont {Grunbaum}}, \ and\ \bibinfo
  {author} {\bibfnamefont {Z.}~\bibnamefont {Ovadyahu}},\ }\href {\doibase
  10.1007/BF00655256} {\bibfield  {journal} {\bibinfo  {journal} {Journal of
  Low Temperature Physics}\ }\textbf {\bibinfo {volume} {10}},\ \bibinfo
  {pages} {231} (\bibinfo {year} {1973}{\natexlab{a}})}\BibitemShut {NoStop}%
\bibitem [{\citenamefont {Cohen}\ and\ \citenamefont
  {Abeles}(1968)}]{Cohen1968}%
  \BibitemOpen
  \bibfield  {author} {\bibinfo {author} {\bibfnamefont {R.~W.}\ \bibnamefont
  {Cohen}}\ and\ \bibinfo {author} {\bibfnamefont {B.}~\bibnamefont {Abeles}},\
  }\href {\doibase 10.1103/PhysRev.168.444} {\bibfield  {journal} {\bibinfo
  {journal} {Phys. Rev.}\ }\textbf {\bibinfo {volume} {168}},\ \bibinfo {pages}
  {444} (\bibinfo {year} {1968})}\BibitemShut {NoStop}%
\bibitem [{\citenamefont {Dynes}\ \emph {et~al.}(1984)\citenamefont {Dynes},
  \citenamefont {Garno}, \citenamefont {Hertel},\ and\ \citenamefont
  {Orlando}}]{Dynes1984}%
  \BibitemOpen
  \bibfield  {author} {\bibinfo {author} {\bibfnamefont {R.~C.}\ \bibnamefont
  {Dynes}}, \bibinfo {author} {\bibfnamefont {J.~P.}\ \bibnamefont {Garno}},
  \bibinfo {author} {\bibfnamefont {G.~B.}\ \bibnamefont {Hertel}}, \ and\
  \bibinfo {author} {\bibfnamefont {T.~P.}\ \bibnamefont {Orlando}},\ }\href
  {\doibase 10.1103/PhysRevLett.53.2437} {\bibfield  {journal} {\bibinfo
  {journal} {Phys. Rev. Lett.}\ }\textbf {\bibinfo {volume} {53}},\ \bibinfo
  {pages} {2437} (\bibinfo {year} {1984})}\BibitemShut {NoStop}%
\bibitem [{\citenamefont {Bachar}\ \emph {et~al.}(2015)\citenamefont {Bachar},
  \citenamefont {Lerer}, \citenamefont {Levy}, \citenamefont {Hacohen-Gourgy},
  \citenamefont {Almog}, \citenamefont {Saadaoui}, \citenamefont {Salman},
  \citenamefont {Morenzoni},\ and\ \citenamefont {Deutscher}}]{Bachar2015}%
  \BibitemOpen
  \bibfield  {author} {\bibinfo {author} {\bibfnamefont {N.}~\bibnamefont
  {Bachar}}, \bibinfo {author} {\bibfnamefont {S.}~\bibnamefont {Lerer}},
  \bibinfo {author} {\bibfnamefont {A.}~\bibnamefont {Levy}}, \bibinfo {author}
  {\bibfnamefont {S.}~\bibnamefont {Hacohen-Gourgy}}, \bibinfo {author}
  {\bibfnamefont {B.}~\bibnamefont {Almog}}, \bibinfo {author} {\bibfnamefont
  {H.}~\bibnamefont {Saadaoui}}, \bibinfo {author} {\bibfnamefont
  {Z.}~\bibnamefont {Salman}}, \bibinfo {author} {\bibfnamefont
  {E.}~\bibnamefont {Morenzoni}}, \ and\ \bibinfo {author} {\bibfnamefont
  {G.}~\bibnamefont {Deutscher}},\ }\href {\doibase 10.1103/PhysRevB.91.041123}
  {\bibfield  {journal} {\bibinfo  {journal} {Physical Review B}\ }\textbf
  {\bibinfo {volume} {91}},\ \bibinfo {pages} {041123} (\bibinfo {year}
  {2015})}\BibitemShut {NoStop}%
\bibitem [{\citenamefont {Bachar}\ \emph {et~al.}(2013)\citenamefont {Bachar},
  \citenamefont {Lerer}, \citenamefont {Hacohen-Gourgy}, \citenamefont
  {Almog},\ and\ \citenamefont {Deutscher}}]{Bachar2013}%
  \BibitemOpen
  \bibfield  {author} {\bibinfo {author} {\bibfnamefont {N.}~\bibnamefont
  {Bachar}}, \bibinfo {author} {\bibfnamefont {S.}~\bibnamefont {Lerer}},
  \bibinfo {author} {\bibfnamefont {S.}~\bibnamefont {Hacohen-Gourgy}},
  \bibinfo {author} {\bibfnamefont {B.}~\bibnamefont {Almog}}, \ and\ \bibinfo
  {author} {\bibfnamefont {G.}~\bibnamefont {Deutscher}},\ }\href {\doibase
  10.1103/PhysRevB.87.214512} {\bibfield  {journal} {\bibinfo  {journal}
  {Physical Review B}\ }\textbf {\bibinfo {volume} {87}},\ \bibinfo {pages}
  {214512} (\bibinfo {year} {2013})}\BibitemShut {NoStop}%
\bibitem [{\citenamefont {Pracht}\ \emph {et~al.}(2016)\citenamefont {Pracht},
  \citenamefont {Bachar}, \citenamefont {Benfatto}, \citenamefont {Deutscher},
  \citenamefont {Farber}, \citenamefont {Dressel},\ and\ \citenamefont
  {Scheffler}}]{Pracht2016}%
  \BibitemOpen
  \bibfield  {author} {\bibinfo {author} {\bibfnamefont {U.~S.}\ \bibnamefont
  {Pracht}}, \bibinfo {author} {\bibfnamefont {N.}~\bibnamefont {Bachar}},
  \bibinfo {author} {\bibfnamefont {L.}~\bibnamefont {Benfatto}}, \bibinfo
  {author} {\bibfnamefont {G.}~\bibnamefont {Deutscher}}, \bibinfo {author}
  {\bibfnamefont {E.}~\bibnamefont {Farber}}, \bibinfo {author} {\bibfnamefont
  {M.}~\bibnamefont {Dressel}}, \ and\ \bibinfo {author} {\bibfnamefont
  {M.}~\bibnamefont {Scheffler}},\ }\href {\doibase 10.1103/PhysRevB.93.100503}
  {\bibfield  {journal} {\bibinfo  {journal} {Physical Review B}\ }\textbf
  {\bibinfo {volume} {93}},\ \bibinfo {pages} {100503} (\bibinfo {year}
  {2016})}\BibitemShut {NoStop}%
\bibitem [{\citenamefont {Pracht}\ \emph {et~al.}(2017)\citenamefont {Pracht},
  \citenamefont {Cea}, \citenamefont {Bachar}, \citenamefont {Deutscher},
  \citenamefont {Farber}, \citenamefont {Dressel}, \citenamefont {Scheffler},
  \citenamefont {Castellani}, \citenamefont {García-García},\ and\
  \citenamefont {Benfatto}}]{Pracht2017}%
  \BibitemOpen
  \bibfield  {author} {\bibinfo {author} {\bibfnamefont {U.~S.}\ \bibnamefont
  {Pracht}}, \bibinfo {author} {\bibfnamefont {T.}~\bibnamefont {Cea}},
  \bibinfo {author} {\bibfnamefont {N.}~\bibnamefont {Bachar}}, \bibinfo
  {author} {\bibfnamefont {G.}~\bibnamefont {Deutscher}}, \bibinfo {author}
  {\bibfnamefont {E.}~\bibnamefont {Farber}}, \bibinfo {author} {\bibfnamefont
  {M.}~\bibnamefont {Dressel}}, \bibinfo {author} {\bibfnamefont
  {M.}~\bibnamefont {Scheffler}}, \bibinfo {author} {\bibfnamefont
  {C.}~\bibnamefont {Castellani}}, \bibinfo {author} {\bibfnamefont {A.~M.}\
  \bibnamefont {García-García}}, \ and\ \bibinfo {author} {\bibfnamefont
  {L.}~\bibnamefont {Benfatto}},\ }\href {\doibase 10.1103/PhysRevB.96.094514}
  {\bibfield  {journal} {\bibinfo  {journal} {Physical Review B}\ }\textbf
  {\bibinfo {volume} {96}},\ \bibinfo {pages} {094514} (\bibinfo {year}
  {2017})}\BibitemShut {NoStop}%
\bibitem [{\citenamefont {Deutscher}\ \emph
  {et~al.}(1973{\natexlab{b}})\citenamefont {Deutscher}, \citenamefont
  {Gershenson}, \citenamefont {Grunbaum},\ and\ \citenamefont
  {Imry}}]{Deutscher1973}%
  \BibitemOpen
  \bibfield  {author} {\bibinfo {author} {\bibfnamefont {G.}~\bibnamefont
  {Deutscher}}, \bibinfo {author} {\bibfnamefont {M.}~\bibnamefont
  {Gershenson}}, \bibinfo {author} {\bibfnamefont {E.}~\bibnamefont
  {Grunbaum}}, \ and\ \bibinfo {author} {\bibfnamefont {Y.}~\bibnamefont
  {Imry}},\ }\href {\doibase 10.1116/1.1318416} {\bibfield  {journal} {\bibinfo
   {journal} {Journal of Vacuum Science and Technology}\ }\textbf {\bibinfo
  {volume} {10}},\ \bibinfo {pages} {697} (\bibinfo {year}
  {1973}{\natexlab{b}})}\BibitemShut {NoStop}%
\bibitem [{\citenamefont {Lerer}\ \emph {et~al.}(2014)\citenamefont {Lerer},
  \citenamefont {Bachar}, \citenamefont {Deutscher},\ and\ \citenamefont
  {Dagan}}]{Lerer2014}%
  \BibitemOpen
  \bibfield  {author} {\bibinfo {author} {\bibfnamefont {S.}~\bibnamefont
  {Lerer}}, \bibinfo {author} {\bibfnamefont {N.}~\bibnamefont {Bachar}},
  \bibinfo {author} {\bibfnamefont {G.}~\bibnamefont {Deutscher}}, \ and\
  \bibinfo {author} {\bibfnamefont {Y.}~\bibnamefont {Dagan}},\ }\href
  {\doibase 10.1103/PhysRevB.90.214521} {\bibfield  {journal} {\bibinfo
  {journal} {Physical Review B}\ }\textbf {\bibinfo {volume} {90}},\ \bibinfo
  {pages} {214521} (\bibinfo {year} {2014})}\BibitemShut {NoStop}%
\bibitem [{\citenamefont {Mondal}\ \emph {et~al.}(2011)\citenamefont {Mondal},
  \citenamefont {Kamlapure}, \citenamefont {Chand}, \citenamefont {Saraswat},
  \citenamefont {Kumar}, \citenamefont {Jesudasan}, \citenamefont {Benfatto},
  \citenamefont {Tripathi},\ and\ \citenamefont {Raychaudhuri}}]{Mondal2011}%
  \BibitemOpen
  \bibfield  {author} {\bibinfo {author} {\bibfnamefont {M.}~\bibnamefont
  {Mondal}}, \bibinfo {author} {\bibfnamefont {A.}~\bibnamefont {Kamlapure}},
  \bibinfo {author} {\bibfnamefont {M.}~\bibnamefont {Chand}}, \bibinfo
  {author} {\bibfnamefont {G.}~\bibnamefont {Saraswat}}, \bibinfo {author}
  {\bibfnamefont {S.}~\bibnamefont {Kumar}}, \bibinfo {author} {\bibfnamefont
  {J.}~\bibnamefont {Jesudasan}}, \bibinfo {author} {\bibfnamefont
  {L.}~\bibnamefont {Benfatto}}, \bibinfo {author} {\bibfnamefont
  {V.}~\bibnamefont {Tripathi}}, \ and\ \bibinfo {author} {\bibfnamefont
  {P.}~\bibnamefont {Raychaudhuri}},\ }\href {\doibase
  10.1103/PhysRevLett.106.047001} {\bibfield  {journal} {\bibinfo  {journal}
  {Phys. Rev. Lett.}\ }\textbf {\bibinfo {volume} {106}},\ \bibinfo {pages}
  {047001} (\bibinfo {year} {2011})}\BibitemShut {NoStop}%
\bibitem [{\citenamefont {Valenti}\ \emph {et~al.}(2019)\citenamefont
  {Valenti}, \citenamefont {Henriques}, \citenamefont {Catelani}, \citenamefont
  {Maleeva}, \citenamefont {Gr\"unhaupt}, \citenamefont {von L\"upke},
  \citenamefont {Skacel}, \citenamefont {Winkel}, \citenamefont {Bilmes},
  \citenamefont {Ustinov}, \citenamefont {Goupy}, \citenamefont {Calvo},
  \citenamefont {Beno\^{\i}t}, \citenamefont {Levy-Bertrand}, \citenamefont
  {Monfardini},\ and\ \citenamefont {Pop}}]{Valenti2019}%
  \BibitemOpen
  \bibfield  {author} {\bibinfo {author} {\bibfnamefont {F.}~\bibnamefont
  {Valenti}}, \bibinfo {author} {\bibfnamefont {F.}~\bibnamefont {Henriques}},
  \bibinfo {author} {\bibfnamefont {G.}~\bibnamefont {Catelani}}, \bibinfo
  {author} {\bibfnamefont {N.}~\bibnamefont {Maleeva}}, \bibinfo {author}
  {\bibfnamefont {L.}~\bibnamefont {Gr\"unhaupt}}, \bibinfo {author}
  {\bibfnamefont {U.}~\bibnamefont {von L\"upke}}, \bibinfo {author}
  {\bibfnamefont {S.~T.}\ \bibnamefont {Skacel}}, \bibinfo {author}
  {\bibfnamefont {P.}~\bibnamefont {Winkel}}, \bibinfo {author} {\bibfnamefont
  {A.}~\bibnamefont {Bilmes}}, \bibinfo {author} {\bibfnamefont {A.~V.}\
  \bibnamefont {Ustinov}}, \bibinfo {author} {\bibfnamefont {J.}~\bibnamefont
  {Goupy}}, \bibinfo {author} {\bibfnamefont {M.}~\bibnamefont {Calvo}},
  \bibinfo {author} {\bibfnamefont {A.}~\bibnamefont {Beno\^{\i}t}}, \bibinfo
  {author} {\bibfnamefont {F.}~\bibnamefont {Levy-Bertrand}}, \bibinfo {author}
  {\bibfnamefont {A.}~\bibnamefont {Monfardini}}, \ and\ \bibinfo {author}
  {\bibfnamefont {I.~M.}\ \bibnamefont {Pop}},\ }\href {\doibase
  10.1103/PhysRevApplied.11.054087} {\bibfield  {journal} {\bibinfo  {journal}
  {Phys. Rev. Applied}\ }\textbf {\bibinfo {volume} {11}},\ \bibinfo {pages}
  {054087} (\bibinfo {year} {2019})}\BibitemShut {NoStop}%
\bibitem [{\citenamefont {Chand}\ \emph {et~al.}(2012)\citenamefont {Chand},
  \citenamefont {Saraswat}, \citenamefont {Kamlapure}, \citenamefont {Mondal},
  \citenamefont {Kumar}, \citenamefont {Jesudasan}, \citenamefont {Bagwe},
  \citenamefont {Benfatto}, \citenamefont {Tripathi},\ and\ \citenamefont
  {Raychaudhuri}}]{Chand2012}%
  \BibitemOpen
  \bibfield  {author} {\bibinfo {author} {\bibfnamefont {M.}~\bibnamefont
  {Chand}}, \bibinfo {author} {\bibfnamefont {G.}~\bibnamefont {Saraswat}},
  \bibinfo {author} {\bibfnamefont {A.}~\bibnamefont {Kamlapure}}, \bibinfo
  {author} {\bibfnamefont {M.}~\bibnamefont {Mondal}}, \bibinfo {author}
  {\bibfnamefont {S.}~\bibnamefont {Kumar}}, \bibinfo {author} {\bibfnamefont
  {J.}~\bibnamefont {Jesudasan}}, \bibinfo {author} {\bibfnamefont
  {V.}~\bibnamefont {Bagwe}}, \bibinfo {author} {\bibfnamefont
  {L.}~\bibnamefont {Benfatto}}, \bibinfo {author} {\bibfnamefont
  {V.}~\bibnamefont {Tripathi}}, \ and\ \bibinfo {author} {\bibfnamefont
  {P.}~\bibnamefont {Raychaudhuri}},\ }\href {\doibase
  10.1103/PhysRevB.85.014508} {\bibfield  {journal} {\bibinfo  {journal} {Phys.
  Rev. B}\ }\textbf {\bibinfo {volume} {85}},\ \bibinfo {pages} {014508}
  (\bibinfo {year} {2012})}\BibitemShut {NoStop}%
\bibitem [{\citenamefont {Sherman}\ \emph {et~al.}(2015)\citenamefont
  {Sherman}, \citenamefont {Pracht}, \citenamefont {Gorshunov}, \citenamefont
  {Poran}, \citenamefont {Jesudasan}, \citenamefont {Chand}, \citenamefont
  {Raychaudhuri}, \citenamefont {Swanson}, \citenamefont {Trivedi},
  \citenamefont {Auerbach}, \citenamefont {Scheffler}, \citenamefont
  {Frydman},\ and\ \citenamefont {Dressel}}]{Sherman2015}%
  \BibitemOpen
  \bibfield  {author} {\bibinfo {author} {\bibfnamefont {D.}~\bibnamefont
  {Sherman}}, \bibinfo {author} {\bibfnamefont {U.~S.}\ \bibnamefont {Pracht}},
  \bibinfo {author} {\bibfnamefont {B.}~\bibnamefont {Gorshunov}}, \bibinfo
  {author} {\bibfnamefont {S.}~\bibnamefont {Poran}}, \bibinfo {author}
  {\bibfnamefont {J.}~\bibnamefont {Jesudasan}}, \bibinfo {author}
  {\bibfnamefont {M.}~\bibnamefont {Chand}}, \bibinfo {author} {\bibfnamefont
  {P.}~\bibnamefont {Raychaudhuri}}, \bibinfo {author} {\bibfnamefont
  {M.}~\bibnamefont {Swanson}}, \bibinfo {author} {\bibfnamefont
  {N.}~\bibnamefont {Trivedi}}, \bibinfo {author} {\bibfnamefont
  {A.}~\bibnamefont {Auerbach}}, \bibinfo {author} {\bibfnamefont
  {M.}~\bibnamefont {Scheffler}}, \bibinfo {author} {\bibfnamefont
  {A.}~\bibnamefont {Frydman}}, \ and\ \bibinfo {author} {\bibfnamefont
  {M.}~\bibnamefont {Dressel}},\ }\href {http://dx.doi.org/10.1038/nphys3227}
  {\bibfield  {journal} {\bibinfo  {journal} {Nature Physics}\ }\textbf
  {\bibinfo {volume} {11}},\ \bibinfo {pages} {188} (\bibinfo {year}
  {2015})}\BibitemShut {NoStop}%
\bibitem [{\citenamefont {Abeles}(1977)}]{Abeles1977}%
  \BibitemOpen
  \bibfield  {author} {\bibinfo {author} {\bibfnamefont {B.}~\bibnamefont
  {Abeles}},\ }\href {\doibase 10.1103/PhysRevB.15.2828} {\bibfield  {journal}
  {\bibinfo  {journal} {Physical Review B}\ }\textbf {\bibinfo {volume} {15}},\
  \bibinfo {pages} {2828} (\bibinfo {year} {1977})}\BibitemShut {NoStop}%
\bibitem [{\citenamefont {Beloborodov}\ \emph {et~al.}(2007)\citenamefont
  {Beloborodov}, \citenamefont {Lopatin}, \citenamefont {Vinokur},\ and\
  \citenamefont {Efetov}}]{Beloborodov2007}%
  \BibitemOpen
  \bibfield  {author} {\bibinfo {author} {\bibfnamefont {I.~S.}\ \bibnamefont
  {Beloborodov}}, \bibinfo {author} {\bibfnamefont {A.~V.}\ \bibnamefont
  {Lopatin}}, \bibinfo {author} {\bibfnamefont {V.~M.}\ \bibnamefont
  {Vinokur}}, \ and\ \bibinfo {author} {\bibfnamefont {K.~B.}\ \bibnamefont
  {Efetov}},\ }\href {\doibase 10.1103/RevModPhys.79.469} {\bibfield  {journal}
  {\bibinfo  {journal} {Rev. Mod. Phys.}\ }\textbf {\bibinfo {volume} {79}},\
  \bibinfo {pages} {469} (\bibinfo {year} {2007})}\BibitemShut {NoStop}%
\bibitem [{\citenamefont {Georges}\ \emph {et~al.}(1996)\citenamefont
  {Georges}, \citenamefont {Kotliar}, \citenamefont {Krauth},\ and\
  \citenamefont {Rozenberg}}]{Georges1996}%
  \BibitemOpen
  \bibfield  {author} {\bibinfo {author} {\bibfnamefont {A.}~\bibnamefont
  {Georges}}, \bibinfo {author} {\bibfnamefont {G.}~\bibnamefont {Kotliar}},
  \bibinfo {author} {\bibfnamefont {W.}~\bibnamefont {Krauth}}, \ and\ \bibinfo
  {author} {\bibfnamefont {M.~J.}\ \bibnamefont {Rozenberg}},\ }\href {\doibase
  10.1103/RevModPhys.68.13} {\bibfield  {journal} {\bibinfo  {journal} {Rev.
  Mod. Phys.}\ }\textbf {\bibinfo {volume} {68}},\ \bibinfo {pages} {13}
  (\bibinfo {year} {1996})}\BibitemShut {NoStop}%
\bibitem [{\citenamefont {Kubo}(1962)}]{Kubo1962}%
  \BibitemOpen
  \bibfield  {author} {\bibinfo {author} {\bibfnamefont {R.}~\bibnamefont
  {Kubo}},\ }\href {\doibase 10.1143/JPSJ.17.975} {\bibfield  {journal}
  {\bibinfo  {journal} {Journal of the Physical Society of Japan}\ }\textbf
  {\bibinfo {volume} {17}},\ \bibinfo {pages} {975} (\bibinfo {year}
  {1962})}\BibitemShut {NoStop}%
\bibitem [{\citenamefont {Gerber}\ \emph {et~al.}(1997)\citenamefont {Gerber},
  \citenamefont {Milner}, \citenamefont {Deutscher}, \citenamefont
  {Karpovsky},\ and\ \citenamefont {Gladkikh}}]{Gerber1997}%
  \BibitemOpen
  \bibfield  {author} {\bibinfo {author} {\bibfnamefont {A.}~\bibnamefont
  {Gerber}}, \bibinfo {author} {\bibfnamefont {A.}~\bibnamefont {Milner}},
  \bibinfo {author} {\bibfnamefont {G.}~\bibnamefont {Deutscher}}, \bibinfo
  {author} {\bibfnamefont {M.}~\bibnamefont {Karpovsky}}, \ and\ \bibinfo
  {author} {\bibfnamefont {A.}~\bibnamefont {Gladkikh}},\ }\href {\doibase
  10.1103/PhysRevLett.78.4277} {\bibfield  {journal} {\bibinfo  {journal}
  {Phys. Rev. Lett.}\ }\textbf {\bibinfo {volume} {78}},\ \bibinfo {pages}
  {4277} (\bibinfo {year} {1997})}\BibitemShut {NoStop}%
\bibitem [{\citenamefont {Bachar}\ \emph {et~al.}(2020)\citenamefont {Bachar},
  \citenamefont {Levy}, \citenamefont {Prokscha}, \citenamefont {Suter},
  \citenamefont {Morenzoni}, \citenamefont {Salman},\ and\ \citenamefont
  {Deutscher}}]{Bachar2020}%
  \BibitemOpen
  \bibfield  {author} {\bibinfo {author} {\bibfnamefont {N.}~\bibnamefont
  {Bachar}}, \bibinfo {author} {\bibfnamefont {A.}~\bibnamefont {Levy}},
  \bibinfo {author} {\bibfnamefont {T.}~\bibnamefont {Prokscha}}, \bibinfo
  {author} {\bibfnamefont {A.}~\bibnamefont {Suter}}, \bibinfo {author}
  {\bibfnamefont {E.}~\bibnamefont {Morenzoni}}, \bibinfo {author}
  {\bibfnamefont {Z.}~\bibnamefont {Salman}}, \ and\ \bibinfo {author}
  {\bibfnamefont {G.}~\bibnamefont {Deutscher}},\ }\href {\doibase
  10.1103/PhysRevB.101.024424} {\bibfield  {journal} {\bibinfo  {journal}
  {Phys. Rev. B}\ }\textbf {\bibinfo {volume} {101}},\ \bibinfo {pages}
  {024424} (\bibinfo {year} {2020})}\BibitemShut {NoStop}%
\bibitem [{\citenamefont {Zhang}\ \emph {et~al.}(2019)\citenamefont {Zhang},
  \citenamefont {Kalashnikov}, \citenamefont {Lu}, \citenamefont {Kamenov},
  \citenamefont {DiNapoli},\ and\ \citenamefont {Gershenson}}]{Zhang2019}%
  \BibitemOpen
  \bibfield  {author} {\bibinfo {author} {\bibfnamefont {W.}~\bibnamefont
  {Zhang}}, \bibinfo {author} {\bibfnamefont {K.}~\bibnamefont {Kalashnikov}},
  \bibinfo {author} {\bibfnamefont {W.-S.}\ \bibnamefont {Lu}}, \bibinfo
  {author} {\bibfnamefont {P.}~\bibnamefont {Kamenov}}, \bibinfo {author}
  {\bibfnamefont {T.}~\bibnamefont {DiNapoli}}, \ and\ \bibinfo {author}
  {\bibfnamefont {M.}~\bibnamefont {Gershenson}},\ }\href {\doibase
  10.1103/PhysRevApplied.11.011003} {\bibfield  {journal} {\bibinfo  {journal}
  {Phys. Rev. Applied}\ }\textbf {\bibinfo {volume} {11}},\ \bibinfo {pages}
  {011003} (\bibinfo {year} {2019})}\BibitemShut {NoStop}%
\bibitem [{\citenamefont {Shearrow}\ \emph {et~al.}(2018)\citenamefont
  {Shearrow}, \citenamefont {Koolstra}, \citenamefont {Whiteley}, \citenamefont
  {Earnest}, \citenamefont {Barry}, \citenamefont {Heremans}, \citenamefont
  {Awschalom}, \citenamefont {Shirokoff},\ and\ \citenamefont
  {Schuster}}]{Shearrow2018}%
  \BibitemOpen
  \bibfield  {author} {\bibinfo {author} {\bibfnamefont {A.}~\bibnamefont
  {Shearrow}}, \bibinfo {author} {\bibfnamefont {G.}~\bibnamefont {Koolstra}},
  \bibinfo {author} {\bibfnamefont {S.~J.}\ \bibnamefont {Whiteley}}, \bibinfo
  {author} {\bibfnamefont {N.}~\bibnamefont {Earnest}}, \bibinfo {author}
  {\bibfnamefont {P.~S.}\ \bibnamefont {Barry}}, \bibinfo {author}
  {\bibfnamefont {F.~J.}\ \bibnamefont {Heremans}}, \bibinfo {author}
  {\bibfnamefont {D.~D.}\ \bibnamefont {Awschalom}}, \bibinfo {author}
  {\bibfnamefont {E.}~\bibnamefont {Shirokoff}}, \ and\ \bibinfo {author}
  {\bibfnamefont {D.~I.}\ \bibnamefont {Schuster}},\ }\href {\doibase
  10.1063/1.5053461} {\bibfield  {journal} {\bibinfo  {journal} {Applied
  Physics Letters}\ }\textbf {\bibinfo {volume} {113}},\ \bibinfo {pages}
  {212601} (\bibinfo {year} {2018})},\ \Eprint
  {http://arxiv.org/abs/https://doi.org/10.1063/1.5053461}
  {https://doi.org/10.1063/1.5053461} \BibitemShut {NoStop}%
\bibitem [{\citenamefont {Cheng}\ \emph {et~al.}(2016)\citenamefont {Cheng},
  \citenamefont {Wu}, \citenamefont {Laurita}, \citenamefont {Singh},
  \citenamefont {Chand}, \citenamefont {Raychaudhuri},\ and\ \citenamefont
  {Armitage}}]{Cheng2016}%
  \BibitemOpen
  \bibfield  {author} {\bibinfo {author} {\bibfnamefont {B.}~\bibnamefont
  {Cheng}}, \bibinfo {author} {\bibfnamefont {L.}~\bibnamefont {Wu}}, \bibinfo
  {author} {\bibfnamefont {N.~J.}\ \bibnamefont {Laurita}}, \bibinfo {author}
  {\bibfnamefont {H.}~\bibnamefont {Singh}}, \bibinfo {author} {\bibfnamefont
  {M.}~\bibnamefont {Chand}}, \bibinfo {author} {\bibfnamefont
  {P.}~\bibnamefont {Raychaudhuri}}, \ and\ \bibinfo {author} {\bibfnamefont
  {N.~P.}\ \bibnamefont {Armitage}},\ }\href {\doibase
  10.1103/PhysRevB.93.180511} {\bibfield  {journal} {\bibinfo  {journal} {Phys.
  Rev. B}\ }\textbf {\bibinfo {volume} {93}},\ \bibinfo {pages} {180511}
  (\bibinfo {year} {2016})}\BibitemShut {NoStop}%
\bibitem [{\citenamefont {Mattis}\ and\ \citenamefont
  {Bardeen}(1958)}]{Mattis1958}%
  \BibitemOpen
  \bibfield  {author} {\bibinfo {author} {\bibfnamefont {D.~C.}\ \bibnamefont
  {Mattis}}\ and\ \bibinfo {author} {\bibfnamefont {J.}~\bibnamefont
  {Bardeen}},\ }\href {\doibase 10.1103/PhysRev.111.412} {\bibfield  {journal}
  {\bibinfo  {journal} {Phys. Rev.}\ }\textbf {\bibinfo {volume} {111}},\
  \bibinfo {pages} {412} (\bibinfo {year} {1958})}\BibitemShut {NoStop}%
\bibitem [{\citenamefont {Fontaine}\ and\ \citenamefont
  {Meunier}(1972)}]{Fontaine1972}%
  \BibitemOpen
  \bibfield  {author} {\bibinfo {author} {\bibfnamefont {A.}~\bibnamefont
  {Fontaine}}\ and\ \bibinfo {author} {\bibfnamefont {F.}~\bibnamefont
  {Meunier}},\ }\href {https://doi.org/10.1007/BF02422516} {\bibfield
  {journal} {\bibinfo  {journal} {Physik der kondensierten Materie}\ }\textbf
  {\bibinfo {volume} {14}},\ \bibinfo {pages} {119} (\bibinfo {year}
  {1972})}\BibitemShut {NoStop}%
\end{thebibliography}

\end{document}